\documentclass[fleqn,usenatbib]{mnras}

\usepackage{newtxtext,newtxmath}

\usepackage[T1]{fontenc}
\usepackage{ae,aecompl}

\usepackage{graphicx}	
\usepackage{amsmath}	

\defcitealias{2021Park}{P21}

\title[Eruptive YSOs in nearby SFRs]{Photometric and Spectroscopic monitoring of YSOs in nearby star forming regions. I. Eruptive YSOs}

\author[C. Contreras Pe\~{n}a et al.]{
Carlos Contreras Pe\~{n}a$^{1}$,\thanks{E-mail: ccontreras@snu.ac.kr (CCP)}
Gregory J. Herczeg$^{2,3}$,
Mizna Ashraf$^{4}$,
Jessy Jose$^{4}$,
\newauthor Ho-Gyu Lee$^{5}$, 
Doug Johnstone$^{6,7}$,
Jeong-Eun Lee$^{1}$,
Xing-yu Zhou$^{2,3}$,
Hanpu Liu$^{2,3}$,
\newauthor  Sung-Yong Yoon$^{5,8}$
\\
$^{1}$ Department of Physics and Astronomy, Seoul National University, 1 Gwanak-ro, Gwanak-gu, Seoul 08826, Republic of Korea\\
$^{2}$ Kavli Institute for Astronomy and Astrophysics, Peking University, Yiheyuan 5, Haidian Qu, 100871 Beijing, People’s Republic of China\\
$^{3}$ Department of Astronomy, Peking University, Yiheyuan 5, Haidian Qu, 100871 Beijing, People’s Republic of China\\
$^{4}$ Indian Institute of Science Education and Research (IISER) Tirupati, Rami Reddy Nagar, Karakambadi Road, Mangalam (P.O.), Tirupati 517 507, India \\
$^{5}$  Korea Astronomy and Space Science Institute, 776, Daedeok-daero, Yuseong-gu, Daejeon, 34055, Republic of Korea\\
$^{6}$ NRC Herzberg Astronomy and Astrophysics, 5071 West Saanich Road, Victoria, BC, V9E 2E7, Canada\\
$^{7}$Department of Physics and Astronomy, University of Victoria, 3800 Finnerty Road, Elliot Building, Victoria, BC, V8P 5C2, Canada\\
$^{8}$ School of Space Research, Kyung Hee University, 1732, Deogyeong-daero, Giheung-gu, Yongin-si, Gyeonggi-do 17104, Republic of Korea\\
}

\date{Accepted XXX. Received YYY; in original form ZZZ}

\pubyear{2023}

\begin{document}
\label{firstpage}
\pagerange{\pageref{firstpage}--\pageref{lastpage}}
\maketitle

\begin{abstract}
Mid-infrared (mid-IR) variability in young stellar objects (YSOs) is driven by several physical mechanisms, which produce a variety of amplitudes and light curve shapes. One of these mechanisms, variable disk accretion is predicted by models of episodic accretion to drive secular variability, including in the mid-IR.  Because the largest accretion bursts are rare, 
adding new objects to the YSO eruptive variable class  aids  our understanding of the episodic accretion phenomenon and its possible impact on stellar and planetary formation. A previous analysis of 6.5 yr of NeoWISE light curves (3–5 $\mu$m) of $\sim$7000 nearby YSOs found an increase in the fraction of variability and variability amplitude for objects at younger stages of evolution. 
To help interpret these light curves,
we have obtained low- and high-resolution near-IR spectra of 78 objects from this sample of YSOs. In this work, we present the analysis of nine nearby YSOs (d$<1$~kpc) that show the characteristics of known classes of eruptive variable YSOs. We find one FUor-like source, one EX Lupi-type object, and six YSOs with mixed characteristics, or V1647 Ori-like objects. The varied characteristics observed in our sample are consistent with recent discoveries of eruptive YSOs. We discuss how a wide range in YSO outburst parameters (central mass, maximum accretion rate during outburst, evolutionary stage and/or instability leading to the outburst) may play a significant role in the observed spectro-photometric properties of YSO outbursts.
\end{abstract}




\section{Introduction}\label{sec:intro}

Young stellar objects (YSOs) display variability over a wide wavelength range, from X-rays to the sub-mm, with a variety of amplitudes and timescales. Dedicated photometric observations at optical, near- to mid-IR and sub-mm wavelengths have been fundamental to understand the physical mechanisms, associated with the stellar photosphere, accretion disk and envelopes of YSOs, that give rise to variability in these systems \citep{1994Herbst,2001Carpenter,2014Cody,2017Contreras_a,2017Venuti,2020Sergison,2021Lee}. In particular, observations at mid-IR wavelengths are 
a powerful tool to evaluate the variability of embedded protostars that are usually too faint at optical and even near-IR wavelengths. The Young Stellar Object VARiability (YSOVAR) program \citep[e.g.][]{c2011Morales,2014Stauffer} and comparisons between Spitzer and Wide-field Infrared Survey Explorer (WISE) mid-IR photometry \citep{2013Scholz,2014Antoniucci,2019Fischer} revealed the complexity of YSO variability at wavelengths between 3 and 5 $\mu$m.

Recently, \citet{2021Park}, hereafter \citetalias{2021Park}, performed a statistical analysis of the mid-IR variability of 7000 YSOs in nearby star forming regions (SFRs). Monitoring from the Neo-WISE W1 (3.4 $\mu$m) and W2 (4.6 $\mu$m) filter observations from 2013 to 2020 allowed  the classification of the many-year variability in these YSOs as either showing long-term trends (or secular variability) and stochastic variability. {\citetalias{2021Park} find an} increase in the fraction of variability towards the younger stages of evolution.  Embedded objects are also found to show more secular variability and with higher amplitudes than evolved YSOs. 

The variability at mid-IR wavelengths can be explained by various mechanisms, including extinction changes due to inhomogeneous mass distributions in the disk or varying disk geometry \citep{c2011Morales,2021Covey}, short-burst variability due to hydrodynamic interactions between stellar surfaces and inner disk edges \citep{2014Stauffer}, and hot or cold spots on the stellar surfaces \citep{2014Cody}. An important mechanism to drive secular variability is likely variable disk accretion rates, as predicted by the episodic accretion model \citep{2020Contreras,2021Park,2022Fischer}.

Distinguishing between the different mechanisms is difficult through light curves alone. Spectroscopic observations are needed to gain a better understanding of the underlying physical mechanisms leading to the mid-IR variability in nearby YSOs seen in the sample of \citetalias{2021Park}. With this goal in mind, we have been conducting a programme to follow-up a sub-sample of YSOs using low and high-resolution near-infrared spectrographs, including Gemini/GNIRS, Gemini/Flamingos-2, Gemini/IGRINS, IRTF/Spex and Palomar/Triplespec observations, in the period between 2020 May and 2022 June.  

In this work, we focus on YSOs with spectra that confirmed that variability is driven by changes in the accretion rate of the system. YSO outbursts driven by large changes in the accretion rate are still a rare class of objects \citep[see e.g.][]{2018Connelley}. The discovery of eruptive variable YSOs  will aid to further understand these class of objects and the phenomenon of episodic accretion. 
In this work we show the spectroscopic follow-up of a sample of nearby YSOs from \citet{2021Park} that show photometric characteristics of eruptive variables in their light curves.

This work is divided as follows. In section 2 we present an overview of the classification of eruptive YSOs. In Section 3 we explain the sample selection and present a summary of the photometric and spectroscopic data used in this work. In Section 4, we  describe the spectroscopic and photometric characteristics of individual sources analysed in this work. In Section 5 we discuss the mechanisms that could  explain the high-amplitude variability in our sample. In Section 6 we argue how the different parameters in young stars can lead to the mixed spectro-photometric characteristics in outbursting YSOs. Finally, in Section 7 we present a summary of our findings.

\section{An overview of episodic accretion}

In the episodic accretion model of star formation, stars gain most of their mass during short-lived episodes of high accretion (reaching as high as $\dot{\mathrm{M}}\simeq10^{-4}$ M$_{\odot}$ yr$^{-1}$) followed by long periods of quiescent low-level accretion \citep{1996Hartmann}. Instabilities in the disc lead to sudden episodes of enhanced accretion. Many models could explain the instabilities that give rise to the outbursts, which include gravitational and magnetorotational instabilities \citep{2009Zhu, 2020Kadam}, thermal viscous instability \citep{1994Bell,1995Bell}, disc fragmentation \citep{2015Vorobyov}, binary interaction \citep{2010Reipurth}, stellar flybys \citep{2019Cuello} and planet-disc interaction \citep{2004Lodato}. 

Episodic accretion can have an effect on stellar and planet formation. It has been invoked  as one of the accretion models that could explain the observed spread in the luminosity of protostars in areas of star formation \citep[see review by][]{2022Fischer}. The intense accretion outbursts can also have a long-lasting impact on the structure of the central star (\citealp[cf. ][]{2017Baraffe} with \citealp{2017Kunimoto}). The long periods spent at low accretion stages may allow the disc to cool sufficiently to fragment, helping in the production low-mass companions \citep{2012Stamatellos}. Accretion outbursts alter the chemistry of protoplanetary discs \citep{2019Artur}, the location of the snowline of various ices \citep{2016Cieza}, and could affect orbital evolution of any planets, if present  \citep{2013Boss,2021Becker}


The class of YSOs that show accretion-related variability, or eruptive variable YSOs, provide observational evidence for episodic accretion \citep{1996Hartmann}. This class of objects is usually divided according to their photometric and spectroscopic characteristics during outburst \citep[see e.g.][]{2018Connelley}. The original criteria was set by \citet{1977Herbig,1989Herbig} based on a handful of YSOs and on observations taken at optical wavelengths. More recently, this classification has been updated to also include characteristics at longer near- to mid-IR wavelengths \citep[see e.g.][]{2012Lorenzetti, 2018Connelley}

\subsection{The classical FUors and EX Lup type objects}

FUors, named after the archetype FU Orionis, show high-amplitude outbursts \citep[$\Delta$V$\sim$6 mag, e.g. ][]{2022Hillenbrandb}, that can last for decades (typically longer than 10 years), due to the sudden increase of the accretion rate which can reach as high as 10$^{-4}$~M$_{\odot}$~yr$^{-1}$ \citep{1996Hartmann}. During outburst, FUors are characterised by an absorption spectrum with very few emission lines. The most common characteristics in FUors are the H$\alpha$ line with a P Cygni profile, Na I D ($\lambda$5900~\AA) absorption, $^{12}$CO bandhead absorption ($\lambda$22935~\AA) and a triangular H-band continuum due to H$_{2}$O absorption ($\lambda$13300~\AA). FUors also show a change in spectral type with observed wavelength going from F-G in the optical to K-M in the near-infrared. During outburst, the accretion disk dominates emission in the system, where absorption lines arise due to a cooler disk surface compared with the viscously-heated midplane. The change in spectral type is due to emission arising at different disk radii \citep[][]{1977Herbig,1996Hartmann,2010Reipurth_b,2018Connelley,2018Hillenbrand,2022Fischer}.

EX Lup outbursts \citep[originally named EXors by e.g.][]{1977Herbig}, are considered as the less dramatic counterparts of FUors. The outbursts in these systems can last a few weeks to several months and reach similar amplitudes to FUor outbursts. In addition, outbursts in these systems have sometimes been seen to be repetitive. The spectra of EX Lup type objects are also different to FUors, as they are dominated by emission lines during maximum light. In the near-IR, Na I ($\lambda$22060~\AA) and $^{12}$CO bandhead ($\lambda$22935~\AA) emission arise from the surface layers of a hot inner disc. These lines go into absorption during the quiescent state. The accepted interpretation is that at low accretion rates, the central star dominates the emission from the system, leading to the photospheric absorption profile. On the other hand, an increase in the value of the accretion rate leads to an increase of UV radiation at the accretion shock, which in turn increases the inner disk temperature and thus favours CO emission \citep{2009Lorenzetti}. In general, EX Lup type objects observed at minimum light were found to be no different than typical T Tauri stars \citep{1989Herbig,2008Herbig,2012Lorenzetti,2014Audard}.

Due to the differences between the two classes, it was the accepted view that these represented distinct types of objects. FUors were a phenomenon related to YSOs at younger evolutionary stages, while EX Lup type objects were characteristic of more evolved Class II YSOs. New discoveries, however, have shown the existence of FUor outbursts during the Class II stage \citep{2019Contreras} and have blurred the two-class system of eruptive YSOs  \citep{2017Contreras,2021Guo}.

\subsection{Recent discoveries}\label{ssec:new}

Multi-epoch, optical \citep[Gaia, ZTF, Pan-STARRS, ASSASN,][]{2016Chambers,2018Bellm,2018Jayasinghe,2021Hodgkin}, near-IR \citep[UKIDSS GPS, VVV, PGIR,][]{c2008Lucas,2012Saito,2016Moore}, mid-IR \citep[Spitzer, WISE, NEOWISE,][]{2003Benjamin,2010Wright,2014Mainzer} and sub-mm \citep[JCMT,][]{2017Herczeg} surveys have led to the discovery of a larger sample of eruptive YSOs. The new discoveries include examples of objects that fit the original classification, i.e. bona-fide FUors \citep[e.g. VVVv721, Gaia17bpi, G286.2032+0.1740, Gaia18dvy, DR4\_V20, PGIR 20cdi,][]{2017Contreras, 2018Hillenbrand,2020Cheng, 2020Szegedi-Elek, 2021Guo, 2021Hillenbrand_a} and EX Lup type objects \citep[e.g Gaia20eae,][]{2022Cruz,2022Ghosh}. The majority of recent discoveries, however, show a mixture of EX Lup/FUor spectroscopic and photometric characteristics. The latter are often classified under different names, such as V1647 Ori-like \citep[][]{2022Fischer}, `peculiar' \citep[][]{2018Connelley} or `MNors' \citep[][]{2017Contreras_a}. 

These objects with mixed characteristics exhibit a wide range in behaviours. OO Ser \citep[also known as a deeply embedded outburst star,][]{1996Hodapp} had a $\sim$10 years-long outburst, but during maximum brightness it showed a featureless, red rising continuum spectrum \citep{2007Kospal}. V1647 Ori, perhaps the archetype of eruptive YSOs with mixed characteristics, shows multiple outbursts with duration longer than EX Lup type objects (i.e. 2 years or longer). The outbursts in V1647 Ori show spectroscopic characteristics of both EX Lup type objects and FUors \citep[see e.g.][]{2022Fischer}. The eruptive YSOs arising from the VVV survey are dominated by objects with emission line spectra and outbursts with duration longer than those of EX Lup type objects. For example, 17 out of 19 eruptive YSOs from the VVV survey presented in \citet{2017Contreras} are classified as V1647 Ori-like objects. These are interpreted as eruptive YSOs where, in spite of the large outbursts, magnetospheric accretion still controls how mass is accreted onto the central star \citep[see][]{2017Contreras_a, 2017Contreras, 2020Guo, 2021Guo}. Similar characteristics are observed in eruptive YSOs ASASSN-13db \citep[][]{2017Sicilia}, Gaia 19bey \citep[][]{2020Hodapp}, ESO-H$\alpha$ 99 \citep{2019Hodapp}, UKIDSS J185318.36$+$012454.5 \citep{2017Niko} and V1318 Cyg \citep{2022Hillenbrand}. There have also been examples of YSOs showing FUor-like spectra at peak brightness, but with outbursts that are much shorter than expected in bona-fide FUors \citep[VVVv322, Gaia 21bty; ][]{2017Contreras, 2021Hillenbrand}.

Periodic outbursts, which do no fit into the original classification and are often classified as peculiar, have been observed in a number of YSOs. V371 Ser \citep[EC53,][]{2012Hodapp, 2017Yoo} shows quasi-periodic variability across a wide wavelength range (1--1100 $\mu$m) with a period of P$\sim$573 d. The variability is likely due to a cyclical buildup and draining of mass in the inner disk \citep[][]{2020YHLee}.
Similar periodic behaviour has been observed in L1634 IRS7 \citep{2015Hodapp}, V347 Aur \citep[][]{2020Dahm}, and 71 VVV sources in \citet{2022Guo}. Dynamical perturbations from stellar or planetary companions have been invoked as a possible driver of the periodic variability \citep{2020YHLee}.
 
 Finally, surveys conducted at mid-IR and sub-mm wavelengths have permitted the detection of outbursts in YSOs at the earlier stages of young stellar evolution (Class 0/I YSOs) and that are generally too faint to be observed at the wavelengths that are commonly used to classify outbursts (optical and near-IR). Classification at these stages is sometimes difficult, with some objects showing nearly featureless spectra similar to OO Ser \citep[V723 Car, 2MASS 22352345$+$7517076,][]{2015Tapia,2019Kun}. Many outburst sources are too faint for spectroscopic follow up even at the brightest point of the outburst, making it difficult to classify these objects into the FUor/EX Lup scheme, as is the case for YSOs NWISE-F J213723.5$+$665145, HOPS 383, HOPS 12 and HOPS 124 \citep{2015Safron, 2020Connelley, 2022Zakri} and many of the YSOs discovered by the JCMT Transient Survey \citep{2021Lee}.

 New discoveries of eruptive variables have raised more questions as to the phenomenon of accretion related variability. It seems that we are not observing just two distinct classes of objects with well defined characteristics but rather events with a continuum of empirical measurements \citep{2022Fischer}. Understanding this continuum requires the discovery and analysis of a larger sample of outbursting YSOs.

\section{Observations}

With the aim of understanding the physical mechanism driving the variability in the sources analysed by \citetalias{2021Park}, we 
have collected spectroscopic data for 78 YSOs from \citetalias{2021Park}. Figure \ref{fig:sample_info} shows the amplitude, standard deviation over fiducial standard deviation (SD/$\sigma$ in flux space)\footnote{This parameter is used by \citetalias{2021Park} to quantify the likelihood that the mid-IR variability is real and not arising from measurement uncertainties.}, and magnitude of the 78 YSOs. The figure shows that YSOs with spectroscopic data tend to have the largest amplitudes within the sample of \citetalias{2021Park}. In addition, we have observed the majority of YSOs with $\Delta W2>2$~mag.


\subsection{YSO sample}

The YSOs presented in this work were selected as eruptive variables based on inspection of their near- to mid-IR light curves as well as their spectroscopic characteristics during follow-up. In this sense, we present only those objects where an interpretation as an eruptive variable was possible. Additional YSOs in the sample of \citetalias{2021Park} have light curves that could be classified as eruptive, however, due to lack of spectroscopic follow-up or difficult interpretation of the spectrum, we do not present them in this work. 

For two YSOs from our sample, HOPS 315 and [LAL96] 213, we support our interpretation using additional 850 $\mu$m data arising from the JCMT Transient Survey \citep{2017Herczeg}. The brightening event of HOPS 315 was first noticed in the sub-mm \citep{2021Lee} and later associated to the mid-IR source, which is classified as non-variable in \citetalias{2021Park}, and was therefore included in our follow-up. 
Sources that are seen to vary at sub-mm wavelengths are continuously cross-matched against the sample of \citetalias{2021Park} by our group. 

 One YSO included in our analysis is not in the sample of \citetalias{2021Park}. V565 Mon was selected based on the similarity of its Gaia colour and magnitude to those expected in FUor objects, and was observed during one of the available nights of spectroscopic follow-up.

\begin{figure*}
	\resizebox{1.9\columnwidth}{!}{\includegraphics[angle=0]{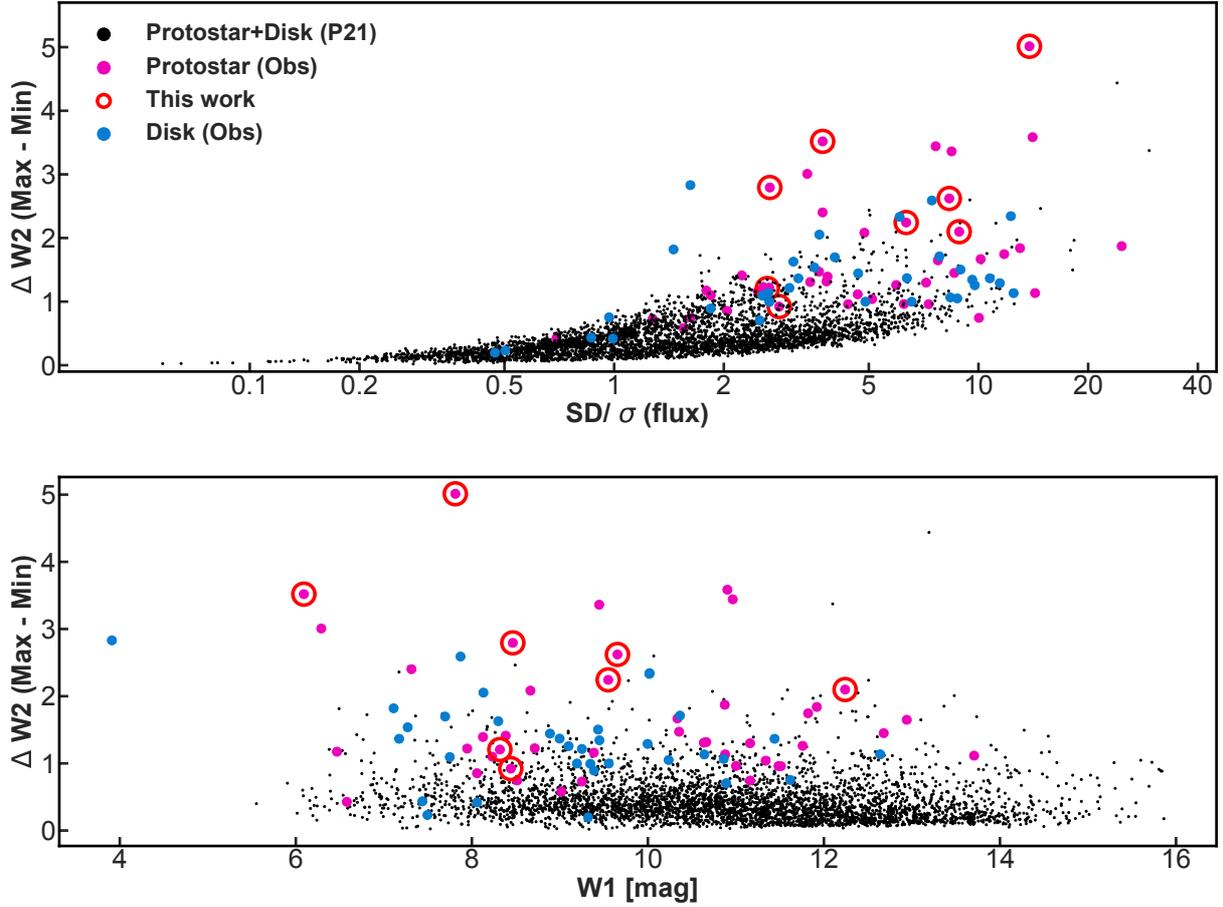}}
	 \caption{$\Delta W2$ vs SD/$\sigma$ (top) and W1 magnitude (bottom) for all of the YSOs analysed in \citetalias{2021Park}. The black circles show YSOs (classified as protostars or disks) where no spectroscopic data has been obtained. YSOs with spectroscopic data are shown in blue (disks) and pink (protostars) circles. Finally YSOs presented in this work as candidate eruptive variables are marked by large red open circles.  } 
    \label{fig:sample_info}
\end{figure*}

Figure \ref{fig:sample_info} shows the properties for eight of the YSOs presented in this work, and that are part of \citetalias{2021Park}. 
These YSOs are preferentially located towards the regions of the diagram with large values of both $\Delta W2$ and SD/$\sigma$.

 In Table \ref{tab:sample} we show the general characteristics of the nine YSOs analysed in this work. The designation of the YSOs from \citetalias{2021Park} and their most common names are shown in columns 1 and 2. The right ascension and declination of the sources are presented in columns 3 and 4. The YSO class, luminosity and distances obtained from previous works are given in columns 5, 6 and 7, respectively. Finally, in columns 8 through 11, we show the $W2$ amplitude and classification of the light curve from the analyses of \citetalias{2021Park} and this work. The differences observed between the amplitudes of columns 8 and 10 arise due to the additional epochs added during the analysis presented in this work (see Section \ref{ssec:photometry}).

The distances presented in Table \ref{tab:sample} are taken from the latest available values for the clouds where the YSOs are located (usually determined from {\it Gaia} DR2 observations). The luminosities presented in this table are also corrected to reflect any changes in the assumed distances. The only exception is for V565 Mon, as there are no {\it Gaia}-based estimates for the dark cloud L1653. \citet{2021Andreasyan} provides a range of distances to the dark cloud that are based on kinematic distances to CO clouds and photometric distances to OB stars \citep{1986Maddalena,2004Kim}. In the case of V565 Mon the distance ($d=1150$~pc) is taken as the average of the distances presented in \citet{2021Andreasyan}. The latter is in agreement with the {\it Gaia} distance estimated for V565 Mon itself , $d=1202$~pc \citep{2021Bailer}.


 \begin{table*}
	\centering
	\caption{YSO Sample Attributes}
	\label{tab:sample}
\resizebox{\textwidth}{!}{
   	\begin{tabular}{lcccccccccc} 
		\hline
		YSO ID \citepalias{2021Park} & Other name & $\alpha$ (J2000) & $\delta$ (J2000) & Class &  Luminosity (L$_{\odot}$)& Distance (pc) & $\Delta W2$ \citepalias{2021Park} & Class \citepalias{2021Park} & $\Delta W2$ (this work) & Class (this work)\\
            \hline
            (1) & (2) & (3) & (4) & (5) & (6) & (7) & (8) & (9) & (10) & (11)\\
		\hline
            D2369$^{a}$ & LDN 1455 IRS3 & 03:28:00.3 & $+$30:08:01.0 & I &  0.36 & 300 & 2.24 & \textit{Irregular} & 2.24 & \textit{Irregular} \\
            M457 & HOPS 267 & 05:41:19.7 & $-$07:50:41.0 & I & 1.1 & 429 &0.79 & \textit{Irregular} & 5.01 & \textit{Burst} \\
	    M713$^{b}$ & HOPS 154 & 05:38:20.1 & $-$06:59:04.9 & I & 0.09 & 389 & 2.15 & \textit{Irregular} & 2.1 & \textit{Irregular} \\
	      D1486$^{c}$ & 2MASS J21013280$+$6811204 & 21:01:32.8	& $+$68:11:20.0 & I  & 6.9 & 341 & 1.98 & \textit{Burst} & 2.62 & \textit{Linear} \\
	      D1826$^{d}$ & 2MASS J21533472$+$4720439 & 21:53:34.7 & $+$47:20:44.0 & F/I & 3.97 & 800 & 0.92 & \textit{Linear} & 0.92 & \textit{Linear} \\
	    M3159 & HOPS 315 & 05:46:03.6 & $-$00:14:49.2 & I & 6.2 & 427 & 0.44& non-variable & 1.21 & non-variable \\
	    D2439$^{e}$ & [LAL96] 213 & 03:29:07.7 & $+$31:21:57.0 & I & 25.9 &  300 & -- & -- & 3.52 & \textit{Curved} \\
	    D1607$^{f}$ & GM Cha & 11:09:28.5 & $-$76:33:28.0 & I & 1.5  & 191  & -- & -- & 2.79 & \textit{Curved} \\
	    -- & V565 Mon & 06:58:02.7 & $-$07:56:43.6 & II & 130 & 1150 & -- & -- & 0.4 & non-variable \\
		\hline	
		\multicolumn{11}{l}{$a$ \citet{2015Dunham} estimates the luminosity as 0.29 L$_{\odot}$ using a distance of 250 pc. The corrected distance is taken from \citet{2018Zucker}}\\
            \multicolumn{11}{l}{$b$ \citet{2016Furlan} estimates the luminosity as 0.1 L$_{\odot}$ using a distance of 420 pc. The corrected distance is taken from \citet{2020Tobin}}\\
		\multicolumn{11}{l}{$c$ \citet{2015Dunham} estimates the luminosity as 3.1 L$_{\odot}$ using a distance of 228 pc. The corrected distance is taken from \citet{2021Szi}}\\
		\multicolumn{11}{l}{$d$ \citet{2015Dunham} estimates the luminosity as 5.6 L$_{\odot}$ using a distance of 950 pc. The corrected distance is taken from \citet{2020Wang}.}\\
		\multicolumn{11}{l}{$e$ \citet{2015Dunham} estimates the luminosity as 18 L$_{\odot}$ using a distance of 950 pc. The corrected distance is taken from \citet{2018Zucker}}\\
            \multicolumn{11}{l}{$e$ \citet{2015Dunham} estimates the luminosity as 0.9 L$_{\odot}$ using a distance of 150 pc. The corrected distance is taken from \citet{2021Galli}}\\
	\end{tabular}}
\end{table*}

\subsection{Photometry}\label{ssec:photometry} 

The original variability analysis from \citetalias{2021Park} is based on mid-IR multi-epoch observations from the {\it WISE} telescope. {\it WISE} surveyed the entire sky in four bands, W1 (3.4 $\mu$m), W2 (4.6 $\mu$m), W3 (12 $\mu$m), and W4 (22 $\mu$m), with the spatial resolutions of 6.1\arcsec, 6.4\arcsec, 6.5\arcsec, and 12\arcsec, respectively, from 2010 January to September \citep{2010Wright}. The survey continued as the NEOWISE Post-Cryogenic Mission, using only the W1 and W2 bands, for an additional 4 months \citep{2011Mainzer}. In 2013 September, WISE was reactivated as the NEOWISE-reactivation mission \citep[NEOWISE-R,][]{2014Mainzer}. NEOWISE-R is still operating and the latest released data set contains observations until mid-December 2021.

The analysis of \citetalias{2021Park} only used 6.5 yr NEOWISE-R data taken between 2013 and 2020. For the photometric analysis presented here, we use the complete WISE/NEOWISE data, covering observations between 2010 until 2021, corresponding to 3 to 4 additional epochs. The data was collected from the NASA/IPAC Infrared Science Archive (IRSA) catalogues, where the queries and averaging of the data was done with the same method as \citetalias{2021Park}. 

In their analysis, \citetalias{2021Park} search for secular and stochastic variability in the long-term data of $\sim$7000 YSOs. Candidate variable stars are selected if they show $\Delta W2/\sigma_{W2}\geq3$. Lomb-Scargle \citep{1976Lomb,1989Scargle} periodogram and linear fits are used to search for secular trends in the light curves of variable objects. YSOs that are well fitted by a linear model are classified as {\it Linear}. YSOs that do not show linear trends, but are well fitted by a sinusoidal light curve are classified as {\it Periodic}, if they have periods of $P<1200$~d, or {\it Curved} if their periods are longer than 1200 d. YSOs that do not fall into the secular variability classes are defined as showing stochastic variability. The latter is further divided into {\it Burst}, {\it Drop} and {\it Irregular} classes. The variability class determined by \citetalias{2021Park} for the YSOs analysed in this work is shown in Table \ref{tab:sample}.

We follow a similar procedure to classify the updated light curves of the YSOs presented in this work. The new classifications are shown in Table \ref{tab:sample}. When searching for secular trends in the data, we follow \citetalias{2021Park} and only use NEOWISE-R data (2013-2021) when using the Lomb-Scargle periodogram, but we use the complete 2010-2021 light curve to fit linear trends. Parameters such as amplitude, average magnitudes and uncertainties are also estimated from the complete light curve. The additional {\it WISE} epochs obtained in 2010 as well as the epochs added during the latest data release explain the observed differences in parameters and classifications found in Table \ref{tab:sample}.

Additional near- and mid-IR photometry was collected for our YSO sample. The data was collected using public catalogues available in Vizier and the NASA/IPAC IRSA. The additional data was only used to visually inspect the longer term ($\sim$20 years) trends and was not used in the determination of parameters, such as amplitude, nor was it included to re-classify the variability of the YSO. 

For two sources, HOPS 315 and [LAL96] 213, we include 850 $\mu$m data arising from the JCMT Transient Survey \citep{2017Herczeg, 2021Lee, 2022Johnstone}. For more details on the observations, data reduction and calibration of the survey see \citet{2017Mairs}.

\subsection{Spectroscopic observations}

The near-IR observations presented here were obtained as part of a larger programme to follow up interesting variables in the sample of \citetalias{2021Park} using low and high resolution spectrographs at Gemini, Palomar and IRTF observatories. 

 Our first proposals requested observations with FLAMINGOS-2 through Gemini partnerships with Republic of Korea and the Dominion of Canada.  After an initial set of observations, we then accessed time on Palomar/Triplespec through the Telescope Access Program in People's Republic of China and NASA IRTF/SpeX through open time offered by the United States of America.  With this range in facilities, we assigned fainter sources to Gemini and brighter sources to IRTF and Palomar.

In the following we describe the data acquisition and reduction for the different instruments used in the observations presented in this work. A summary is presented in Table \ref{tab:spec}.

\subsubsection{Flamingos-2}

We obtained K-band spectra of GM Cha on 22 February 2021 using Flamingos-2 \citep{elkenberry04} on the 8.1 m Gemini South telescope at Cerro Pachon, Chile (programme GS-2021A-Q-110, PI Johnstone). We used the R3K grism with a 3-pxiel wide long-slit to achieve a resolution of $R=1600$. We obtained $12\times30$s exposures nodding along the slit in an ABBA pattern. Observations of B9V star HD 96092 were also acquired in the same setup for a telluric correction. All reduction steps were executed with standard procedures using the GEMINI package in IRAF.

\subsubsection{SpeX}

We obtained near-IR spectra of the YSOs LDN 1455 IRS3, HOPS267, HOPS 154, 2MASS J21533472$+$4720439, HOPS 315 and $[$LAL96$]$ 213 on 23-27 November 2021 with SpeX \citep{rayner03} mounted at the NASA Infrared Telescope Facility (IRTF) on Mauna Kea (programme 2021B043, PI Herczeg).  The cross-dispersed spectra cover 0.7--2.5 $\mu$m  spectra at $R\sim 2000$, obtained with  the 0.3\arcsec and 0.5\arcsec slits. Total integration times ranged from 1800 to 3600s, with individual exposures of 300s.  Bright A0V standard stars were observed for telluric calibration. All spectra were reduced and calibrated using Spextool version 4.1 \citep{cushing04}. 

\subsubsection{TripleSpec}

We obtained 1--2.4 $\mu$m observations ($R=2700$) of 2MASS J21013280$+$6811204 and V565 Mon on 12 December 2021 using Triplespec \citep{herter08} mounted on the Palomar 200-inch Hale telescope (PI Herczeg).  We obtained 4$\times$30s and 4$\times$300s exposures, nodding along the list on an ABBA pattern, for V565 Mon and 2MASS J21013280$+$6811204, respectively. Bright A0V standard stars were observed for telluric calibration. All spectra were reduced and calibrated using Spextool version 4.1 \citep{cushing04}.     

 \begin{table}
	\centering
	\caption{Spectroscopic observations presented in this work}
	\label{tab:spec}
\resizebox{\columnwidth}{!}{
   	\begin{tabular}{lccc} 
		\hline
		YSO & Instrument & $R$ & Date\\
		\hline
            LDN 1455 IRS3 & IRTF/Spex & 2000 & 27-Nov-2021\\
            HOPS 267 & IRTF/Spex & 2000 & 27-Nov-2021 \\
	    HOPS 154 & IRTF/Spex & 2000 & 27-Nov-2021\\
	    2MASS J21013280$+$6811204 & Palomar/TripleSpec & 2700 & 12-Dec-2021 \\
	    2MASS J21533472$+$4720439 & IRTF/Spex & 2000 & 23-Nov-2021\\
	    HOPS 315 & IRTF/Spex & 2000 & 23-Nov-2021 \\
	    $[$LAL96$]$ 213 & IRTF/Spex & 2000 & 24-Nov-2021 \\
	    GM Cha & Gemini/Flamingos 2 & 1600 & 22-Feb-2021   \\
	    V565 Mon & Palomar/TripleSpec & 2700 & 12-Dec-2021 \\
		\hline
	\end{tabular}}
\end{table}

\section{The YSOs}\label{sec:main}

In this section we discuss the various characteristics of individual sources, including information from the literature, photometric behaviour and classifications, and the observed spectroscopic characteristics. 
These characteristics are only briefly analysed, as it is the purpose of this work to place these sources into the broader context of the classification of eruptive variability.

\textbf{\textit{LDN 1455 IRS3}}: Located in the LDN 1455 dark cloud and a member of the Perseus complex, at a distance of 300 pc \citep{2018Zucker}. This Class I YSO \citep{2012Kryukova} had already been catalogued as a candidate EX Lup type object based only on its high-amplitude variability between {\it Spitzer} and {\it WISE} observations \citep{2014Antoniucci}. Both \citetalias{2021Park} and this work detect high-amplitude variability ($\Delta$W2$=2.24$~mag) that result in the source being classified as \textit{Irregular} (see Table \ref{tab:sample}). The light-curve classification, as defined in \citetalias{2021Park}, determines how well the variation can be fitted by either a sinusoidal or linear model. The light curve of the YSO cannot be fitted by these specific models, hence the \textit{Irregular} classification. However the light curve does show indications of some periodicity in the brightening events of the source.

\begin{figure*}
	\resizebox{\columnwidth}{!}{\includegraphics[angle=0]{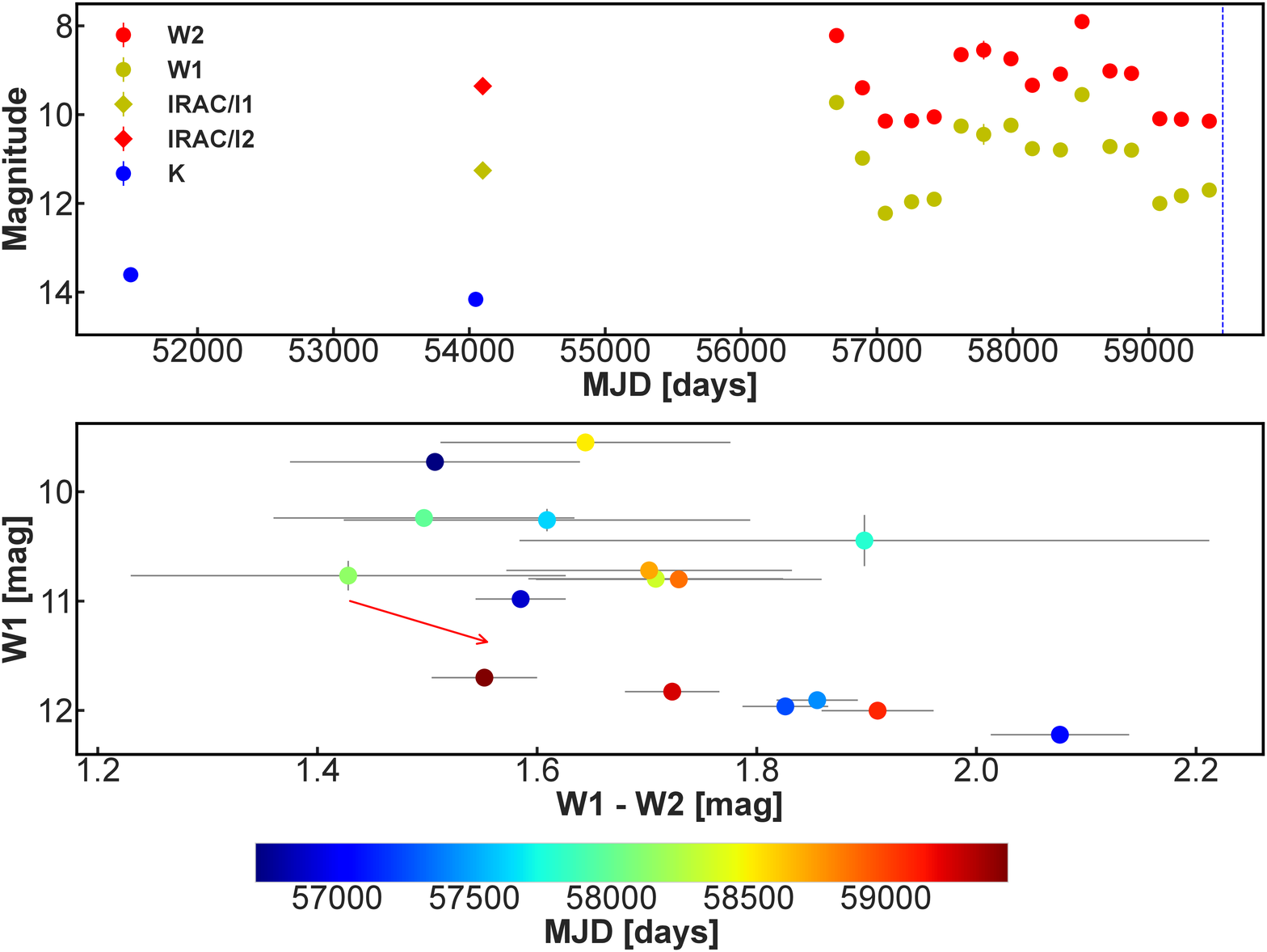}}
	\resizebox{\columnwidth}{!}{\includegraphics[angle=0]{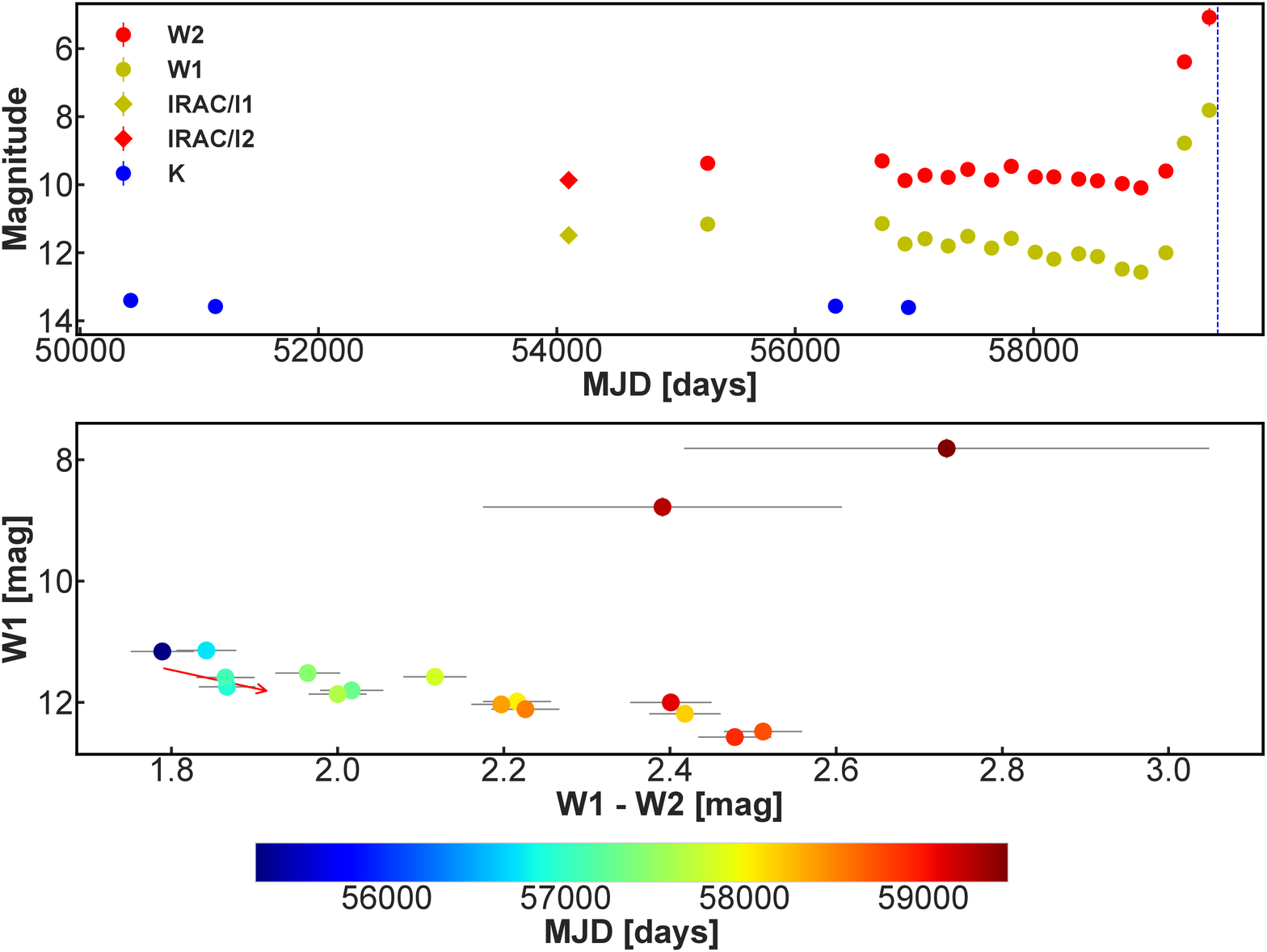}}\\
	\resizebox{\columnwidth}{!}{\includegraphics[angle=0]{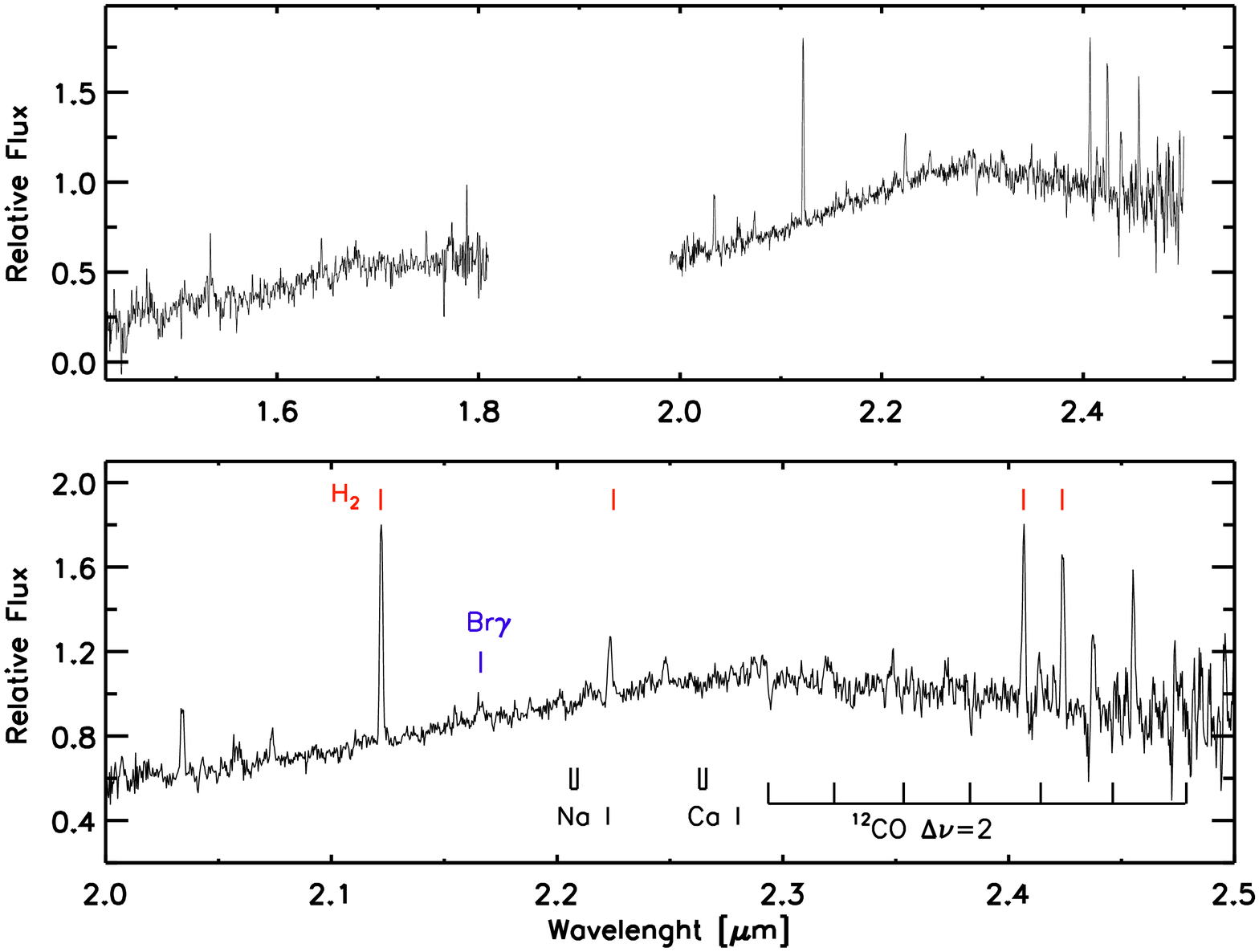}}
	\resizebox{\columnwidth}{!}{\includegraphics[angle=0]{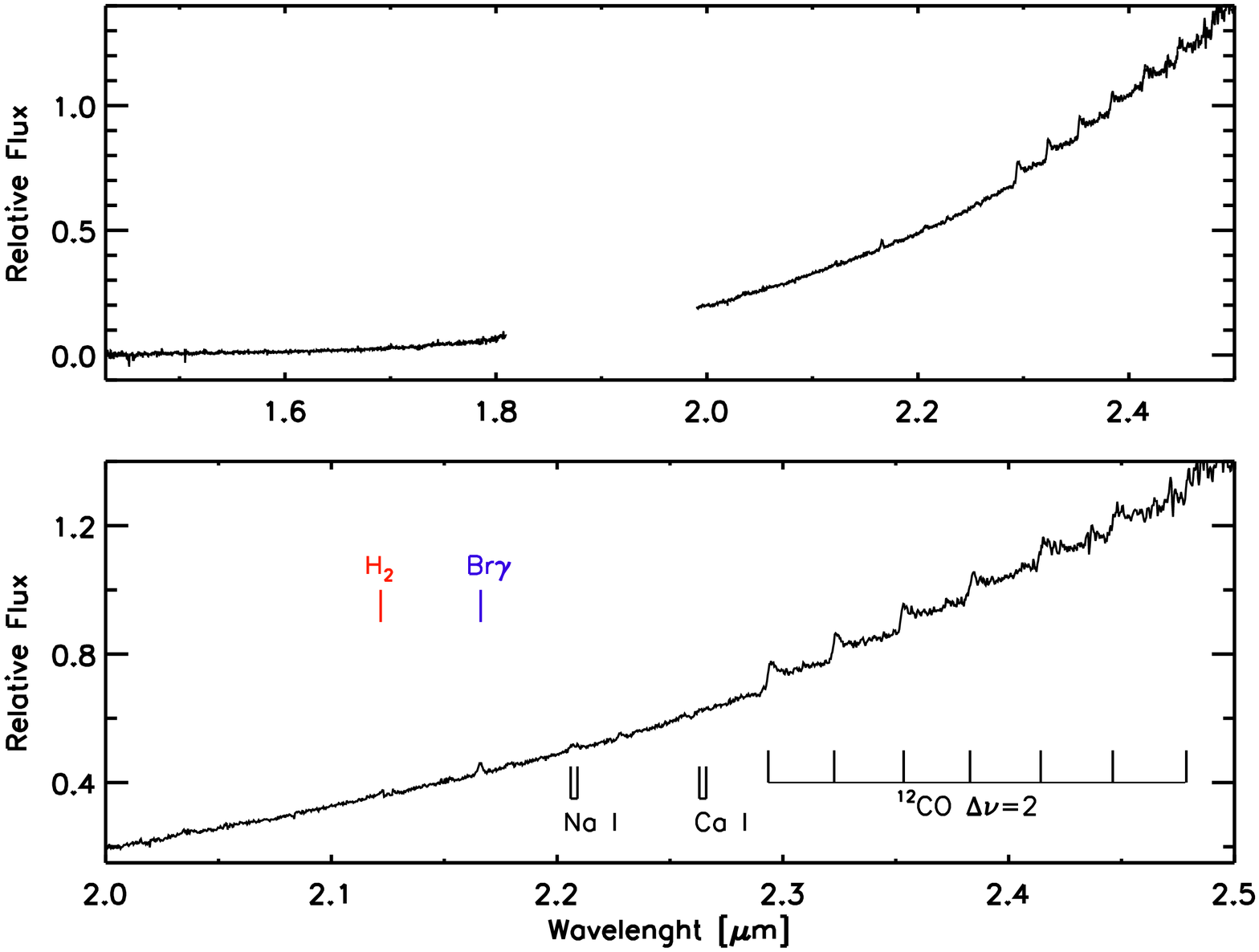}}
	 \caption{(top) K-band (blue), 3.4 $\mu$m (yellow) and 4.6 $\mu$m (red) light curve of LDN 1455 IRS3 (left) and HOPS 267 (right). The date of spectroscopic observations is marked by a dashed blue line. (middle) $W1$ vs $W1-W2$ colour-magnitude diagram for LDN 1455 IRS3 (left) and HOPS 267 (right). The plots only include data obtained by the {\it WISE} telescope. The red arrow marks the reddening line for A$_{V}=20$~mag, using the extinction law of \citet{2019Wang}. (bottom) IRTF/Spex spectrum of LDN 1455 IRS3 (left) and HOPS 267 (right). The upper panel shows the H and K spectra of the sources. In the bottom panel, only the K-band portion of the spectrum is shown, along with the location of typical emission/absorption features in YSOs.}
    \label{fig:LDN1455}
\end{figure*}

The IRTF spectrum (Fig. \ref{fig:LDN1455}) shows strong H$_{2}$ emission, as well as CO and H$_{2}$O absorption (beyond 2.29 $\mu$m), although only the $^{12}$CO $\nu=2-0$ band appears to be detected. Br$\gamma$ emission is apparent but weak. The faint H-band continuum shows some hints of triangular shape due to the H$_{2}$O absorption.
The closest photometric point to the epoch of spectroscopic observations shows the YSO at minimum brightness. However, if the object follows the same pattern of past variability, it is very likely that the YSO was observed close to maximum brightness.

\textbf{\textit{HOPS 267}}: This YSO is located in the Orion Nebula Cluster at a distance of 429 pc \citep{2020Tobin}. It was first classified as a Class I YSO by \citet{2012Megeath} and is part of The Herschel Orion Protostar Survey \citep[HOPS,][]{2016Furlan}, with an SED that is well fit by a YSO model with $L_{\rm bol}=1.1~L_{\odot}$.

Figure \ref{fig:LDN1455} shows that the YSO displays irregular low amplitude variability up to the epochs analysed by \citetalias{2021Park}. This variability seems to follow the  interstellar reddening line. However, the latest epochs in NEOWISE show a large amplitude ($\Delta$W2=5.01 mag) increase, with the source becoming much redder in the process. The high-amplitude variability in the last epochs  changes its classification to the \textit{Burst} class.

We obtained spectroscopic observations apparently as the system was still getting brighter. The object shows a red rising spectrum with marked Br-$\gamma$ and CO emission.

\textbf{\textit{HOPS 154}}: This YSO is located in the Orion Nebula Cluster at a distance of 389 pc \citep{2020Tobin}. It was first detected and classified as a protostar by \citet{2012Megeath} based on 2MASS and {\it Spitzer} observations. The YSO is part of the HOPS survey, with a SED that is well fit by a model with a bolometric luminosity of $L_{\rm bol}=0.09~L_{\odot}$. \citet{2012Kryukova} classifies the object as a Class I YSO based on its $2-24 \mu$m spectral index of $\alpha=0.55$, although the large variability of this object (see Fig. \ref{fig:hops154}) likely affects the spectral slope $\alpha$. Nevertheless, the bolometric temperature and luminosity of the system estimated by \citet{2016Furlan} agree with a Class I SED classification.

The near-to mid-IR light curve of HOPS 154 (Fig. \ref{fig:hops154}) shows irregular, high-amplitude variability. The YSO is classified by \citetalias{2021Park} as an \textit{Irregular} variable star with an amplitude of $\Delta W2=2.15$~mag. The additional epochs analysed in this work do not change the classification nor the amplitude of this system. The $W1$ vs $W1-W2$ colour-magnitude diagram of Fig. \ref{fig:hops154} shows that the YSO gets systematically bluer as it changes into the bright state. This behaviour appears to repeat throughout the different outbursts of the YSO.

The $JHK$ IRTF spectrum, obtained when the YSO was close to its brightest magnitude, shows a H-band continuum with a triangular shape due to H$_{2}$O absoprtion, and strong CO bandhead absorption beyond 2.29 $\mu$m (Fig. \ref{fig:hops154} ).

\textbf{\textit{2MASS J21013280$+$6811204}}: This source is a Class I YSO located in the Lynds 1172/1174 dark cloud at a distance of 341 pc \citep{2021Szi}. 

The mid-IR light curve of the YSO is classified as \textit{Linear} due to the recent outburst (starting close to MJD$=$58000 d) with $\Delta W2=2.66$~mag. The evolution of the W1 vs W1$-$W2 curve during the outburst (Fig. \ref{fig:4990}) does not follow the reddening line, therefore we can discard changes in the extinction along the line of sight as the driving mechanism for the variability of the YSO.

The Palomar/TripleSpec spectrum of the YSO, taken close to maximum, shows a triangular H-band spectrum, $^{12}$CO bandhead absorption, and no emission from atomic Hydrogen lines. H$_{2}$ emission, likely associated to an outflow, can be seen in the spectrum of the source.

\textbf{\textit{2MASS J21533472$+$4720439}}: This object is part of the YSO sample of \citet{2015Dunham} and is located in the IC5146 molecular cloud at a distance of 800 pc \citep{2020Wang}. The YSO has an infrared spectral index $\alpha$ that ranges between 0.24 \citep{2009Gutermuth} and 0.4 \citep{2015Dunham}, giving a classification as either a flat-spectrum or Class I YSO.

The mid-IR light curve of the YSO (Fig. \ref{fig:4990}) is categorised as \textit{Linear} with and amplitude of 0.92 mag in W2. Both the classification and amplitude remain the same between the analysis by \citetalias{2021Park} and this work.
The IRTF/Spex spectrum (\ref{fig:4990}) appears as a relatively featureless red rising spectrum. Only Br $\gamma$ and $^{12}$CO bandhead emission are apparent.

\textbf{\textit{HOPS 315}}: The YSO \citep[also commonly known as HH26 IRS, see e.g.][]{1997Davis} is located in the L1630 molecular cloud at a distance of 427 pc \citep{2020Tobin}. The SED of the source has a spectral index, $\alpha=2$, consistent with a Class I classification.  HOPS 315 is also the exciting source of the large scale molecular outflow HH26A/C/D \citep{1997Davis,2008Antoniucci}. 

The mid-IR light curve shows a change in amplitude between the analysis by \citetalias{2021Park} and our work, going from 0.42 to 1.21 mag in W2. In both cases the light curve does not pass the criteria established by \citetalias{2021Park} to classify the YSO into any of the variability classes. However, this object is included as the mid-IR brightness increase from the latest 2 epochs of NEOWISE is also observed by the JCMT transient survey \citep{2021Lee} at 850 $\mu$m (Fig. \ref{fig:hops315}). The correlated variability between mid-IR and sub-mm wavelengths is likely related to changes in the accretion rate, with the sub-mm flux variability arising as the envelope temperature increases as a response to the rise in the accretion luminosity \citep[see e.g.][]{2020Contreras}. The most recent observations of HOPS 315 at sub-mm wavelengths also shows it rising more steeply (Mairs et al.\ in preparation).  We may be catching the initial stages of the recent outburst.

Previous observations of HOPS 315 show a spectrum dominated by emission lines of H$_{2}$, Br$\gamma$ and $^{12}$CO bandhead \citep{2004Simon,2008Antoniucci,2011Davis}. The IRTF spectrum observed in this work (Fig.~\ref{fig:hops315}) shows the same characteristics.

\textbf{\textit{[LAL96] 213}}: This source was originally classified as a YSO by \citet{1996Lada} in their study of the NGC1333 region. Located at a distance of 300 pc \citep{2018Zucker}, the YSO shows a rising SED and has been categorised as a Class I YSO by several authors \citep[see e.g.][]{2009Evans,2009Gutermuth,2010Connelley,2015Dunham}.

This YSO has the second largest variability amplitude within our sample with $\Delta W2=3.52$~mag and also shows correlated variability with the sub-mm observations of the Gould Belt Survey \citep{mairs17b,2007Ward} and the JCMT transient survey (see Fig. \ref{fig:hops315}). This correlated variability between the mid-IR and sub-mm has already been noticed by the work of \citet{2020Contreras}, and it is very likely due to variable accretion in the system.

Previous near-IR spectra of the source has been collected in 2006 and 2019 by \citet{2010Connelley} and \citet{2021Fiorellino} respectively. The former observations detect H$_{2}$, Na I, Br$\gamma$ and weak $^{12}$CO emission, whilst in the KMOS spectrum from \citet{2021Fiorellino} Br$\gamma$ and H$_{2}$ emission are present, but there are no obvious features at the wavelengths of Na I or $^{12}$CO. Fig. \ref{fig:hops315} shows that those observations appear to have taken place at different brightness levels. Our observation, taken close to the lowest point in the light curve, shows similarities to the spectrum of \citet{2021Fiorellino}, with no obvious features of Na I or $^{12}$CO. The changes between the \citet{2010Connelley} spectrum and those of \citet{2021Fiorellino} and our work are likely due to the large magnitude changes between the epochs of observations. The W1 vs W1$-$W2 CMD shows the variability strongly departs from the expected changes due to variable extinction along the line of sight.

\textbf{\textit{GM Cha}}: This source is a Class I YSO located in the Chamaleon I dark cloud \citep[distance of 191 pc,][]{2017Long, 2021Galli}. It has been previously classified as an eruptive variable by \citet{2007Persi}. The classification arises due to an observed $\Delta K$=2 mag brightening over a period of 3 years. \citet{2007Persi} find similarities to known FUor/EX Lup outbursts. However, the accretion rate at maximum brightness measured by \citet{2007Persi} is $\simeq 10^{-7}$~M$_{\odot}$~yr$^{-1}$, a value that places this YSO into the EX Lup class according to the authors.

The mid-IR light curve of the YSO (Fig. \ref{fig:hops154}) is classified as {\it Curved} as it is well fitted by a sinusoidal function, albeit with a long period of $P=4800$~ days. On top of this long-term behaviour the YSO displayed a 2.8 mag outburst that lasted for about 1 year. The figure also shows the near-IR K-band photometry of the source, which includes the data published by \citet{2007Persi}. The changes observed by \citet{2007Persi} at these wavelengths do not seem related to a high-amplitude, short-term burst, but rather seem to follow the long-term {\it Curved} trend. 

The mid-IR variability of GM Cha, both the long-term trend and the short-term outburst, are likely being driven by changes in the accretion rate, as the variation does not follow the reddening line in Fig. \ref{fig:hops154}. The 1 year outburst probably lead to much higher values of the accretion rate than those found by \citet{2007Persi}.

The K-band spectrum of the source (Fig. \ref{fig:hops154}) is dominated by emission from ro-vibrational transitions of H$_{2}$, probably related to the YSO outflow. A spectrum of GM Cha presented in \citet{2003Gomez} shows similar features as the spectrum presented in this work. From Fig. \ref{fig:hops154} 
the observations of \citet{2003Gomez} appear to have been taken at a similar brightness level as our observations. Given this, the similarity of the characteristics between the two spectra is not surprising. 

Unfortunately our spectroscopic observations were taken after the 2.8 mag outburst has subsided. Therefore we are unable to observe the expected characteristics of outbursting YSOs at high accretion levels (e.g. $^{12}$CO in emission or absorption). This points to the difficulties in classifying eruptive variable YSOs when these are not observed at maximum brightness (see Section \ref{sec:discussion}).

\textbf{\textit{V565 Mon}}: This YSO is the illuminating source of the Parsamian 17 nebula, located in the dark cloud LDN 1653 at a distance of 1120 pc \citep{2021Andreasyan}. The system is considered to be the driver of the Herbig-Haro outflow HH 947 A/B \citep{2008Magakian}. The SED of the YSO is  consistent with that of a source with $L_{bol}=130 L_{\odot}$ \citep[see][]{2021Andreasyan}. Based on its {\it WISE} colours, \citet{2016Fischer} classifies this source as a Class II YSO. 

V565 Mon has been the subject of previous optical spectroscopic observations, where \citet{2021Andreasyan} finds a late F to early G-type spectrum with strong double-peaked H$\alpha$ emission that is divided by a narrow absorption component. In addition the YSO shows Na I Doublet and Ba II absorption. Li I ($\lambda$6707 \AA) absorption, indicating the youth of the system, is also observed. Given the characteristics of the optical spectrum along with the high luminosity of the source, \citet{2021Andreasyan} suggests a possible FUor classification. However, the lack of an H$\alpha$ P Cygni profile in the system and the presence of wide photometric absorption features, argue against the FUor classification, according to \citet{2021Andreasyan}.

The mid-IR light curve of the source (Fig. \ref{fig:V565}) is classified as non-variable and there is no obvious trend in the data. $V$, $r$ and $i$ data obtained from the ASAS-SN catalogue of variable stars \citep{2018Jayasinghe} and the Bochum Galactic Disk Survey \citep{2015Hackstein}, show non-periodic irregular variability with an amplitude of $\simeq$ 0.9 mag. Inspection of digitised photographic plate images from 1953 observations, acquired from the SuperCosmos Sky Survey \citep[SSS][]{2001Hambly}, show V565 Mon at a similar brightness as recent observations from i.e. Pan-STARRS. If the object is truly an FUor, then the beginning of the outburst must have occurred before 1953.

The TripleSpec JHK spectrum obtained in our programme (Fig. \ref{fig:V565}) shows H$_{2}$O absorption, an H-band triangular shape continuum, and $^{12}$CO absorption beyond 2.29$\mu$m. [Fe II] forbidden emission at 1.64 $\mu$m is clearly detected in the system. There is also weak Br$\gamma$ emission, which appears double-peaked, but cannot be confirmed at this resolution.

\section{Nearby eruptive YSOs}

In the continuum light curve analysis of shorter wavelength data (optical, near-IR), one of the main problems when classifying high-amplitude variable YSOs into the eruptive variable class, is that large amplitudes are also expected from other physical mechanisms \citep[see e.g][]{2017Contreras_a, 2018Cody, 2022Hillenbrandc}. In the mid-IR, the majority of YSOs vary with amplitudes that are lower than 1 magnitude \citep{c2011Morales,2021Park}. Larger changes are consistent with variable accretion \citep{2013Scholz}. However, our interpretation of mid-IR colours and amplitudes can be complicated by the effects of molecular emission in W2 \citep{2022Yoon}. In addition, there are rare types of variable stars that could show similarly high amplitudes and perhaps mimic the light curves of eruptive variable YSOs. In the following we discuss some of these variability mechanisms to determine if they could explain the temporal nature of the sources in our sample.

\subsection{Mechanisms driving mid-IR variability in YSOs} \label{sec:sand}

Occultations of a binary system by a circumbinary ring lead to periodic, with P$<200$~d, high-amplitude, $\Delta K$ reaching up to $3$~mag, achromatic variability \citep{2020Garcia,2022Wei}. The most famous member of this class is V582 Mon \citep[KH15D,][]{2018Aronow}. As the brightness changes are not wavelength dependent, mid-IR variability can also reach these high-amplitudes. However, this type of variability is extremely rare \citep{2022Wei} and the short periods, as compared with the cadence of WISE observations, make it unlikely that this form of variability would be mistaken for an outbursting YSO.

Obscuration by structural inhomogeneities in the circumstellar disk also lead to variability in YSOs. The timescale of the variability depends on the location of the structure in the disk, whilst its amplitude is effectively limitless and wavelength dependent \citep{2001Carpenter,2013Hillenbrand, 2017Contreras_a, 2018Guo}. Due to this physical mechanism the colour-magnitude variability of the YSO should move along the reddening line, with objects becoming redder as they become fainter. 

For example, a warped inner disk and clumps of dust and gas at the inner edge of the circumstellar disk, are responsible for the extinction variability in dippers (also known as AA Tau-like systems) and UX ori-type stars, respectively. The timescale for the variability is on the order of days to weeks and with amplitudes that are no larger than 3 magnitudes in the optical \citep{2003Dullemond,2015Stauffer}. The group of dippers in the Orion Nebula Cluster \citep{c2011Morales} show maximum amplitudes of 0.6 magnitudes at 3 $\mu$m. Long-term fading events, which could be associated with occultation by inhomogeneities located at larger distances in the disk, have been observed in e.g. AA Tau\footnote{\citet{2021Covey} interpret the changes as an increase in the scale height of the inner disk.} and RW Aur \citep{2013Bouvier,2016Bozhinova}.  The difference in visual extinction in these events, however, is not large enough to explain the high-amplitudes observed in our sample.

Changes in the scale height of the inner disk can obscure the central object leading to optical fading, but at the same time increase the brightness of the system at mid-IR wavelengths  \citep{2012Flaherty, 2021Covey}. However, this type of variability in the mid-IR has only been observed in a handful of YSOs \citep{2019Bryan} with amplitudes that do not surpass 1 magnitude at 3 $\mu$m \citep[e.g.][]{2021Covey}. We note that radiative transfer models of changes in the scale height predict amplitudes that could reach 1.5 mag, but this only for extreme increases in scale height and for large inclinations of the disk \citep{2019Bryan}. It is therefore unlikely that we are observing such changes in our sample, where $\Delta W2>1.5$~mag for all but three sources.

Nevertheless, in the analysis of the sources in Section \ref{sec:main}, we see that the colour-magnitude variability in 5 out of 9 objects follows the interstellar reddening line (see Table \ref{tab:sample2}). Of these, V565 Mon has strong reasons to be classified as an eruptive YSO (discussed later in this section), and as such is not included in the following discussion. Assuming that variable extinction is driving the variability, and using the amplitudes of Table \ref{tab:sample}, we have estimated the expected maximum difference in extinction, $\Delta$A$_{V}$. Following a standard extinction law \citep[e.g.,][]{2019Wang}, the maximum values of changes in A$_{V}$ would range between 35 and 192 mag, with a mean value of $\Delta$A$_{V}=93$~mag. 


The extremely high values of extinction variability determined above are rare, but have been observed for V2492 Cyg. This is an eruptive variable Class I YSO that shows a complex light curve behaviour. It went into outburst somewhere between {\it Spitzer} and {\it WISE} observations, and has shown quasi-periodic variations with a timescale of $\simeq$221 days since 2010 \citep{2013Hillenbrand}. The near-IR colour changes of the quasi-periodic variations are in agreement with a change in the extinction along the line of sight with a maximum value of A$_{V}=35$~mag. If a dusty structure is responsible for the periodic variability, and assuming that is in a rotating disk around a $\sim1$M$_{\odot}$ star, then this structure would be located at $\sim0.7$~au from the central star \citep{2013Hillenbrand}. The maximum value of A$_{V}$ estimated for the quasi-periodic variation of V2492 Cyg is similar to the lower-end of A$_{V}$ values derived for our sample.

\begin{figure}
	\resizebox{\columnwidth}{!}{\includegraphics[angle=0]{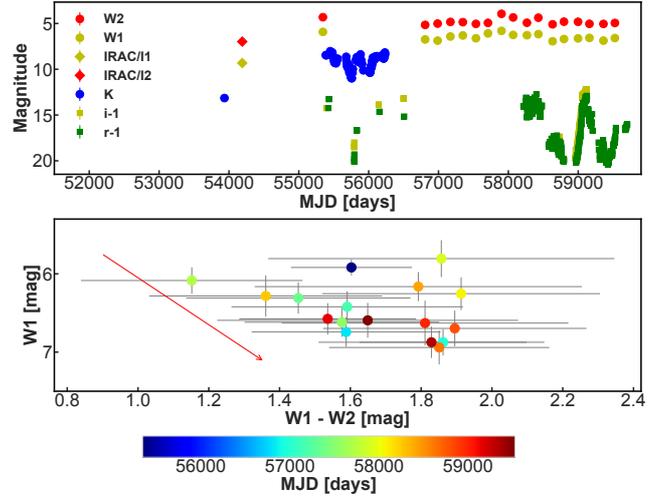}}
	 \caption{Light curve of V2492 Cyg. The red arrow marks the reddening line with A$_{V}=35$~mag.}
    \label{fig:v2492}
\end{figure}

To determine if this type of extinction-related variability could be contaminating our sample of eruptive YSOs, we studied the mid-IR variability of V2492 Cyg over the 2010 to 2021 period (shown in Fig. \ref{fig:v2492}) following the same analysis done for the YSOs in our sample. The statistical analysis shows a mid-IR light curve with $\Delta$W1=1.1 and $\Delta$W2=1.2 mag. Despite the high amplitude, V2492 Cyg does not fulfill the criteria from \citetalias{2021Park} to be classified as a variable object. This is likely due to the high uncertainties in the magnitudes of the source as it lies in the saturated region of {\it WISE}. However, the mid-IR amplitudes do agree with the expectations from variable extinction. 


It is unlikely that an object displaying similar quasi-periodic, extinction-related variations (as displayed by V2492 Cyg after its outburst) would have been selected as a candidate eruptive variable star. The WISE/NEOWISE light curve of V2492 Cyg itself is not classified into any of the variable classes following \citetalias{2021Park} and even inspection of the WISE/NEOWISE light curve would have not lead to a candidate eruptive YSO classification.

Variable extinction, changes in the structure of the inner disk or obscuration by a circumbinary ring could drive some of the variability in the overall sample of YSOs from \citetalias{2021Park}. We cannot completely discard that these processes may play a role in the mid-IR variability of the YSOs presented in this work (especially variable extinction), however, given the arguments presented above it seems unlikely that these mechanisms are the main drivers of the brightness changes in the YSO sample presented in this paper.


\subsection{Variable accretion}

The amplitudes of the variability for 7 out of 9 YSOs presented in this work 
agree
with the expected amplitudes for accretion related outburst in the mid-IR \citep[$\Delta >1-1.5$~mag, see e.g.][]{2013Scholz,2022Hillenbrandb,2022Liu}, with two exceptions: V565 Mon and 2MASS J21533472$+$4720439. 

2MASS J21533472$+$4720439 is perhaps the most difficult source to classify as an eruptive variable. The colour variability seems to follow the reddening line in Fig. \ref{fig:4990}. If so, the maximum value of extinction, A$_{V}=35$~mag, estimated for the source is similar to that of V2492 Cyg discussed above, so an extinction clearing event for this YSO cannot be ruled out completely. The observed spectrum, however, shows CO in emission, consistent with a YSO accreting at high rates. In addition, the shape of the mid-IR light curve is similar to the initial stages of the outburst of Gaia 17bpi \citep{2018Hillenbrand}. In the latter, the mid-IR 
outburst started 1.5-2 years before the onset of the optical outburst. Such behaviour is expected in outbursts that start at greater distances in the disk and propagate inwards \citep[see e.g. Figure 21 in][]{2022Liu}. Whether the mid-IR brightening of 2MASS J21533472$+$4720439 marks the initial stages of an outside-in outburst will only be revealed by continuous monitoring of the source. Given these arguments we classify the YSO as a candidate eruptive variable and propose further monitoring of the source over the next few years in order to confirm the eruptive status of the source.

V565 Mon does not show strong variability but has supporting evidence to be classified as a FUor. It is the illuminating source of a reflection nebula and it has a large bolometric luminosity. The near-IR spectrum of Fig. \ref{fig:V565} is consistent with an M spectral type, whilst \citet{2021Andreasyan} estimate a late F to early G type for the optical spectrum. The change of spectral type between optical and near-infrared wavelengths is one of the defining characteristics of FUor objects. There are, however, some peculiarities in its spectrum that defy the standard FUor classification. \citet{2021Andreasyan} found that the lack of a P Cygni profile for the H$\alpha$ line, and the widths of the photospheric absorption lines were inconsistent with FUor classification. The near-infrared spectrum of Fig.~\ref{fig:V565} shows absorption from low energy rotational transitions of the $v=2-0$ line but lacks features from higher vibrational levels.
As there is no documented outburst, the source does not fulfill all of the criteria to be classified as a  bona-fide FUor, and thus this source is classified as a FUor-like object.


\begin{table*}
	\centering
	\caption{YSO Sample Classification}
	\label{tab:sample2}
\resizebox{\textwidth}{!}{
   	\begin{tabular}{lcccccccc} 
		\hline
		YSO ID \citepalias{2021Park} & Other name & Light Curve & Colour & $\Delta$t (outburst) & Observations & Spectral class & Photometric Class & Final Class \\
		\hline
            D2369 & LDN 1455 IRS3 & \textit{Irregular} & bluer when brighter& $\sim$4.5 yr & Bright State$^{a}$ & FUor & V1647 Ori & V1647 Ori\\
            M457 & HOPS 267  & \textit{Burst} & redder when brighter & $\geq$1 yr & Rising & EX Lup & V1647 Ori & V1647 Ori\\
	    M713 & HOPS 154 & \textit{Irregular} & bluer when brighter & $\sim$1 yr & Bright State & FUor & EX Lup & V1647 Ori\\
	    D1486 & 2MASS J21013280$+$6811204  & \textit{Linear} & redder when brighter & $\geq$3.5 yr  & Rising & FUor & V1647 Ori & V1647 Ori\\
	    D1826$^{b}$ & 2MASS J21533472$+$4720439  & \textit{Linear} & bluer when brighter & $\geq$5 yr & Bright state & EX Lup & V1647 Ori? & Outburst Candidate\\
	    M3159 & HOPS 315 & non-variable & bluer when brighter & $\geq$1 yr & Rising  & EX Lup & V1647 Ori & V1647 Ori\\
	    D2439 & [LAL96] 213 & \textit{Curved} & redder when brighter & $5-7$~yr & quiescence & EX Lup$^{b}$ & V1647 Ori & V1647 Ori \\
	    D1607 & GM Cha & \textit{Curved} & redder when brighter & $<$1.5 yr & quiescence & -- & EX Lup & EX Lup\\
	    -- & V565 Mon & non-variable & bluer when brighter & $>70$ yr & Bright state & FUor & none & FUor-like \\
		\hline
		\multicolumn{9}{l}{$a$ See description in Section \ref{sec:main}}\\
		\multicolumn{9}{l}{$b$ Amplitude and colour behaviour might not be consistent with an eruptive YSO classification (Sections \ref{sec:main} and \ref{sec:sand}).}\\
	\end{tabular}}
\end{table*}

The spectra of HOPS 154, LDN 1455 IRS3 and 2MASS J21013280$+$6811204 show spectroscopic characteristics of bona-fide FUors. In all cases the spectra were taken close to maximum light. However, the light curves of these sources show different characteristics. HOPS 154 and LDN 1455 IRS3 show irregular high-amplitude outbursts with duration of between 1 and 4.5 years, intermediate between EX Lup type and FUor outbursts. The outburst of 2MASS J21013280$+$6811204 is still ongoing and, even though its timescale is now longer than those of EX Lup type outbursts, it cannot be immediately placed on the FUor category, as longer monitoring is needed. 



In the case of HOPS 267, the epochs obtained from the latest NEOWISE data release show a $\Delta W2=5$~mag outburst. The spectrum of HOPS 267, obtained close to maximum in the light curve, shows strong emission lines from Br$\gamma$ and CO. These are usual spectral characteristics of EX Lup type objects in outburst. The duration of the outburst is still unknown, but it looks longer than at least one year, which would likely place it into an intermediate category between EX Lup type and FUor outbursts. This source requires continued monitoring.

HOPS315, [LAL96] 213 and GM Cha, all show EX Lup type spectra, but with various photometric characteristics and/or previous studies that support variable accretion as the driving mechanism of variability. 

A summary on the different spectroscopic and photometric characteristics of our sample is provided in Table \ref{tab:sample2}. In the table we also show a classification based on light curves or spectra alone, as well as the final classification given by the combined characteristics. The sample of nearby YSOs shows a variety of classifications, with V1647 Ori type sources dominating the sample.

\section{The parameter space of outbursting YSOs}\label{sec:discussion}

Recent discoveries of eruptive variables are often challenging to classify, but the underlying physical mechanism remains the same, variable accretion \citep{2022Fischer}. It is possible that the timing of observations and physical parameters of outbursting YSOs, such as central mass, evolutionary stage and/or the instability leading to the outburst, result in the varied spectro-photometric characteristics of eruptive variable YSOs discussed in Section \ref{ssec:new} and the sample of nearby YSOs presented in this work.

One of the primary difficulties in classifying eruptive YSOs is timing. In our sample, GM Cha shows an outburst with an amplitude and colour variation that strongly suggests variable accretion as the main driver. Unfortunately, by the time we were able to follow-up this source, the outburst had subsided and we did not detect any evidence of high accretion rates in the spectrum, making it difficult to place this object into any of the eruptive variable classes. Similar problems are found for WISEA J142238.82$-$611553.7 \citep{2020Lucas} or Gaia 19fct \citep{2022Park}.

Recently, \citet{2022Liu} explored the parameter space of FUor-like outbursts to understand the effect of the mass of the central star (for the range $0.1<M_{\ast}<3.0$\,M$_{\odot}$) and mass accretion rate (for the range $10^{-8}<\dot{M}<10^{-4}$~M$_{\odot}$~yr$^{-1}$) on the observed spectral energy distributions and spectra of these outbursting YSOs.
The models show that as the outburst reaches a determined value of $\dot{M}$, the viscously-heated accretion disk dominates the emission from the system. At this point, the typical spectroscopic characteristics of bona-fide FUors appear, i.e., a triangular shape of the H-band continuum and strong $^{12}$CO absorption at 2.3$\mu$m. Importantly, the value of $\dot{M}$ at which the viscously-heated disk dominates emission depends sensitively on $M_{\ast}$. For example, the viscous disk dominates emission from a $M_{\ast}=0.3 M_{\odot}$ YSO for accretion rates with $\dot{M}\geq4\times10^{-7}$~M$_{\odot}$~yr$^{-1}$. For a more massive YSO, with $M_{\ast}=2.4$M$_{\odot}$, the viscous disk dominates emission when $\dot{M}\geq2\times10^{-6}$~M$_{\odot}$~yr$^{-1}$.

In the analysis by \citet{2022Liu}, when an outburst reaches an accretion rate of $\dot{M}\geq10^{-5}$~M$_{\odot}$~yr$^{-1}$, we should observe a FUor spectrum, regardless of the mass of the central star. It is in this region where the so-called bona-fide FUors probably reside. At values of $10^{-7}<\dot{M}<10^{-5}$~M$_{\odot}$~yr$^{-1}$, 
YSOs could be in a transition region between the point where the fluxes of the disk and the stellar photosphere are equal, and the point where the viscous disk sufficiently outshines the photosphere \citep[see figure 14 in ][]{2022Liu}. The near-IR spectroscopic characteristics of YSO outbursts (specifically $^{12}$CO) falling in this transition region have contributions of both the stellar photosphere and the viscous disk. Although not included by \citet{2022Liu}, a larger stellar irradiation due to an increase in $\dot{M}$ can lead to $^{12}$CO emission from the upper layers of the  accretion disk \citep[see e.g.][]{1991Calvet}. This would also influence the observed characteristics of the $^{12}$CO band.

The above shows that the mass of the central star and the value of the accretion rate play an important role in the observed spectroscopic characteristics and therefore the classification of eruptive YSOs. This may help us understand the characteristics of some of the V1647 Ori-like sources.   

The effect of various sources contributing to the 2.3$\mu$m flux could explain the somewhat weak $^{12}$CO absorption observed in LDN 1455 IRS3, although continuous monitoring will be necessary to confirm the change in the $^{12}$CO line with the brightness of the source. In addition, it is interesting to see that the YSOs in our sample with the higher luminosities (see Table \ref{tab:sample}) show spectra with the $^{12}$CO bandhead in emission. Perhaps these are more massive YSOs where accretion rates have not reached sufficiently high values for viscous accretion to dominate the system. However, this conclusion is affected by the fact that luminosities and spectra were measured at different points in the light curve.

The instability leading to the outburst also plays a role, especially in the shape of the light curve. Instabilities driven by a stellar/planetary companion lead to periodic light curves, which have been observed in several eruptive variable YSOs \citep{2015Hodapp, 2017Yoo, 2020Dahm, 2022Guo}. This type of light curve has generally been classified as peculiar, as they do not fit the classical definition of FUor/EX Lup type outbursts.
In our case, the light curves of HOPS 154 and LDN 1455 IRS3 both show high-amplitude quasi-periodic variability (classified as \textit{Irregular} in our analysis). Perhaps the observed light curves might be due to an unseen companion in the YSOs.

The picture also gets complicated for YSOs that are still deeply embedded in their nascent envelopes. Radiative transfer modelling of the embedded YSO EC53 \citep{2020Baek} shows that parameters such as envelope radius or cavity opening angle affect greatly the observed flux at mid-IR wavelengths. This implies that the amplitudes and light curve shapes of outbursting embedded YSOs could strongly depend on the geometry of their surrounding envelopes. These complicated geometries can affect our ability to analyse the outbursts at the youngest stages of evolution \citep{2020Lucas,2022Yoon}. Sub-mm observations of the light echo shape and delay due to the reprocessing of an accretion burst by the envelope, however, may provide additional useful information \citep{Johnstone13,22Francis}.

The majority of sources in our sample are classified as Class I YSOs and are too faint to be observed at optical wavelengths. It is likely that the youth of these systems is also affecting the observed characteristics of the sample.

\section{Summary}\label{sec:sum}

We have analysed the mid-IR light curves and near-IR spectra of eight YSOs from the sample of \citetalias{2021Park} that are likely new additions to the eruptive variable class. An additional YSO, V565 Mon, that is selected from {\it Gaia}$+${\it WISE} analysis was also presented in this work.

We discuss physical mechanisms that drive mid-IR variability in YSOs, however, we find that the amplitudes of the mid-IR light curves in our sample are more consistent with variable accretion. Nevertheless, we cannot completely discard that changes in the extinction along the line of sight could play a role on the observed variability.

Based on the spectro-photometric characteristics of our sample, such as outburst duration, $^{12}$CO in emission/absorption and/or triangular H-band continuum, we attempt to classify the sources into the more commonly known sub-classes of eruptive variables, i.e. EX Lupi-type or FUors. None of the YSOs in our sample can be classified as a bona-fide FUor, and only in one object falls into the EX Lup class. V565 Mon can only be classified as FUor-like due to the absence of an observed outburst.

In the majority of cases, the outbursts show a mixture of characteristics between those of EX Lupi-type and FUor outbursts. These type of objects, usually classified as V1647 Ori-type outbursts, are the most common class among the recent discoveries of eruptive variable YSOs.

We argue that the varied characteristics in our sample, and in general among eruptive variable YSOs, might be driven by the wide range in the parameters of YSO systems. The mass of the central star, the maximum accretion rate reached during the outburst, the evolutionary stage of the YSO and/or the instability leading to an outburst, can all play a significant role in the observed spectroscopic and photometric characteristics of outbursting YSOs.

We will continue to monitor the nine YSOs presented in this work, as well as the YSOs analysed in \citet{2021Park} using new data releases from mid-IR (NEOWISE) and optical (ZTF) surveys. Eruptive YSOs will be more thoroughly analysed in subsequent works, including modelling of the spectra taking into account the different parameters discussed above. 

In addition the discovery of eruptive YSOs located within 1 kpc, provides ideal candidates for further follow-up using for example the Atacama Large Millimeter/submillimeter Array (ALMA). The sudden rise in luminosity due to an outburst quickly expands the water snow line to larger radii, allowing the observation of complex organic molecules (COMs) in the disk of these YSOs \citep[e.g. ][]{2019Lee}. The nearby outbursting YSOs will be proposed for ALMA observations to study COMs in the early stages of young stellar evolution.

\section*{Acknowledgements}

This work was supported by the National Research Foundation of Korea (NRF) grant funded by the Korean government (MSIT) (grant number 2021R1A2C1011718).  GJH, XY, and HL are supported by grant 12173003 from the National Natural Science Foundation of China. JJ and MA acknowledge the financial support received through the DST-SERB grant SPG/2021/003850. DJ is supported by NRC Canada and by an NSERC Discovery Grant.

This work was in part supported by K-GMT Science Program (GS-2021A-Q-110) of Korea Astronomy and Space Science Institute (KASI).
Based on observations obtained at the international Gemini Observatory, a program of NSF’s NOIRLab, which is managed by the Association of Universities for Research in Astronomy (AURA) under a cooperative agreement with the National Science Foundation on behalf of the Gemini Observatory partnership: the National Science Foundation (United States), National Research Council (Canada), Agencia Nacional de Investigaci\'{o}n y Desarrollo (Chile), Ministerio de Ciencia, Tecnolog\'{i}a e Innovaci\'{o}n (Argentina), Minist\'{e}rio da Ci\^{e}ncia, Tecnologia, Inova\c{c}\~{o}es e Comunica\c{c}\~{o}es (Brazil), and Korea Astronomy and Space Science Institute (Republic of Korea).

This research has made use of the NASA/IPAC Infrared Science Archive, which is funded by the National Aeronautics and Space Administration and operated by the California Institute of Technology. 

This research has made use of the VizieR catalogue access tool, CDS, Strasbourg, France (DOI : 10.26093/cds/vizier). The original description of the VizieR service was published in 2000, A\&AS 143, 23.



\section*{Data Availability}

The data underlying this article are available in the article and in its online supplementary material.

The spectra for various sources will be shared on reasonable request to the corresponding author.



\bibliographystyle{mnras}
\bibliography{spectroscopic_follow_up.bib} 

\begin{thebibliography}{}
\makeatletter
\relax
\def\mn@urlcharsother{\let\do\@makeother \do\$\do\&\do\#\do\^\do\_\do\%\do\~}
\def\mn@doi{\begingroup\mn@urlcharsother \@ifnextchar [ {\mn@doi@}
  {\mn@doi@[]}}
\def\mn@doi@[#1]#2{\def\@tempa{#1}\ifx\@tempa\@empty \href
  {http://dx.doi.org/#2} {doi:#2}\else \href {http://dx.doi.org/#2} {#1}\fi
  \endgroup}
\def\mn@eprint#1#2{\mn@eprint@#1:#2::\@nil}
\def\mn@eprint@arXiv#1{\href {http://arxiv.org/abs/#1} {{\tt arXiv:#1}}}
\def\mn@eprint@dblp#1{\href {http://dblp.uni-trier.de/rec/bibtex/#1.xml}
  {dblp:#1}}
\def\mn@eprint@#1:#2:#3:#4\@nil{\def\@tempa {#1}\def\@tempb {#2}\def\@tempc
  {#3}\ifx \@tempc \@empty \let \@tempc \@tempb \let \@tempb \@tempa \fi \ifx
  \@tempb \@empty \def\@tempb {arXiv}\fi \@ifundefined
  {mn@eprint@\@tempb}{\@tempb:\@tempc}{\expandafter \expandafter \csname
  mn@eprint@\@tempb\endcsname \expandafter{\@tempc}}}

\bibitem[\protect\citeauthoryear{{Andreasyan}}{{Andreasyan}}{2021}]{2021Andreasyan}
{Andreasyan} H.,  2021, \mn@doi [Research in Astronomy and Astrophysics]
  {10.1088/1674-4527/21/3/064}, \href
  {https://ui.adsabs.harvard.edu/abs/2021RAA....21...64A} {21, 064}

\bibitem[\protect\citeauthoryear{{Antoniucci}, {Nisini}, {Giannini}  \&
  {Lorenzetti}}{{Antoniucci} et~al.}{2008}]{2008Antoniucci}
{Antoniucci} S.,  {Nisini} B.,  {Giannini} T.,   {Lorenzetti} D.,  2008,
  \mn@doi [\aap] {10.1051/0004-6361:20077468}, \href
  {https://ui.adsabs.harvard.edu/abs/2008A&A...479..503A} {479, 503}

\bibitem[\protect\citeauthoryear{{Antoniucci}, {Giannini}, {Li Causi}  \&
  {Lorenzetti}}{{Antoniucci} et~al.}{2014}]{2014Antoniucci}
{Antoniucci} S.,  {Giannini} T.,  {Li Causi} G.,   {Lorenzetti} D.,  2014,
  \mn@doi [\apj] {10.1088/0004-637X/782/1/51}, \href
  {https://ui.adsabs.harvard.edu/abs/2014ApJ...782...51A} {782, 51}

\bibitem[\protect\citeauthoryear{{Aronow}, {Herbst}, {Hughes}, {Wilner}  \&
  {Winn}}{{Aronow} et~al.}{2018}]{2018Aronow}
{Aronow} R.~A.,  {Herbst} W.,  {Hughes} A.~M.,  {Wilner} D.~J.,   {Winn} J.~N.,
   2018, \mn@doi [\aj] {10.3847/1538-3881/aa9ed7}, \href
  {https://ui.adsabs.harvard.edu/\#abs/2018AJ....155...47A} {155, 47}

\bibitem[\protect\citeauthoryear{{Artur de la Villarmois}, {J{\o}rgensen},
  {Kristensen}, {Bergin}, {Harsono}, {Sakai}, {van Dishoeck}  \&
  {Yamamoto}}{{Artur de la Villarmois} et~al.}{2019}]{2019Artur}
{Artur de la Villarmois} E.,  {J{\o}rgensen} J.~K.,  {Kristensen} L.~E.,
  {Bergin} E.~A.,  {Harsono} D.,  {Sakai} N.,  {van Dishoeck} E.~F.,
  {Yamamoto} S.,  2019, \mn@doi [\aap] {10.1051/0004-6361/201834877}, \href
  {https://ui.adsabs.harvard.edu/abs/2019A&A...626A..71A} {626, A71}

\bibitem[\protect\citeauthoryear{{Audard} et~al.,}{{Audard}
  et~al.}{2014}]{2014Audard}
{Audard} M.,  et~al., 2014, \mn@doi [Protostars and Planets VI]
  {10.2458/azu_uapress_9780816531240-ch017}, \href
  {http://adsabs.harvard.edu/abs/2014prpl.conf..387A} {pp 387--410}

\bibitem[\protect\citeauthoryear{{Baek} et~al.,}{{Baek}
  et~al.}{2020}]{2020Baek}
{Baek} G.,  et~al., 2020, \mn@doi [\apj] {10.3847/1538-4357/ab8ad4}, \href
  {https://ui.adsabs.harvard.edu/abs/2020ApJ...895...27B} {895, 27}

\bibitem[\protect\citeauthoryear{{Bailer-Jones}, {Rybizki}, {Fouesneau},
  {Demleitner}  \& {Andrae}}{{Bailer-Jones} et~al.}{2021}]{2021Bailer}
{Bailer-Jones} C.~A.~L.,  {Rybizki} J.,  {Fouesneau} M.,  {Demleitner} M.,
  {Andrae} R.,  2021, \mn@doi [\aj] {10.3847/1538-3881/abd806}, \href
  {https://ui.adsabs.harvard.edu/abs/2021AJ....161..147B} {161, 147}

\bibitem[\protect\citeauthoryear{{Baraffe}, {Elbakyan}, {Vorobyov}  \&
  {Chabrier}}{{Baraffe} et~al.}{2017}]{2017Baraffe}
{Baraffe} I.,  {Elbakyan} V.~G.,  {Vorobyov} E.~I.,   {Chabrier} G.,  2017,
  \mn@doi [\aap] {10.1051/0004-6361/201629303}, \href
  {https://ui.adsabs.harvard.edu/abs/2017A&A...597A..19B} {597, A19}

\bibitem[\protect\citeauthoryear{{Becker}, {Batygin}  \& {Adams}}{{Becker}
  et~al.}{2021}]{2021Becker}
{Becker} J.~C.,  {Batygin} K.,   {Adams} F.~C.,  2021, \mn@doi [\apj]
  {10.3847/1538-4357/ac111e}, \href
  {https://ui.adsabs.harvard.edu/abs/2021ApJ...919...76B} {919, 76}

\bibitem[\protect\citeauthoryear{{Bell} \& {Lin}}{{Bell} \&
  {Lin}}{1994}]{1994Bell}
{Bell} K.~R.,  {Lin} D.~N.~C.,  1994, \mn@doi [\apj] {10.1086/174206}, \href
  {https://ui.adsabs.harvard.edu/abs/1994ApJ...427..987B} {427, 987}

\bibitem[\protect\citeauthoryear{{Bell}, {Lin}, {Hartmann}  \& {Kenyon}}{{Bell}
  et~al.}{1995}]{1995Bell}
{Bell} K.~R.,  {Lin} D.~N.~C.,  {Hartmann} L.~W.,   {Kenyon} S.~J.,  1995,
  \mn@doi [\apj] {10.1086/175612}, \href
  {https://ui.adsabs.harvard.edu/abs/1995ApJ...444..376B} {444, 376}

\bibitem[\protect\citeauthoryear{Bellm et~al.,}{Bellm et~al.}{2018}]{2018Bellm}
Bellm E.~C.,  et~al., 2018, \mn@doi [Publications of the Astronomical Society
  of the Pacific] {10.1088/1538-3873/aaecbe}, 131, 018002

\bibitem[\protect\citeauthoryear{{Benjamin} et~al.,}{{Benjamin}
  et~al.}{2003}]{2003Benjamin}
{Benjamin} R.~A.,  et~al., 2003, \mn@doi [\pasp] {10.1086/376696}, \href
  {https://ui.adsabs.harvard.edu/abs/2003PASP..115..953B} {115, 953}

\bibitem[\protect\citeauthoryear{{Boss}}{{Boss}}{2013}]{2013Boss}
{Boss} A.~P.,  2013, \mn@doi [\apj] {10.1088/0004-637X/764/2/194}, \href
  {http://adsabs.harvard.edu/abs/2013ApJ...764..194B} {764, 194}

\bibitem[\protect\citeauthoryear{{Bouvier}, {Grankin}, {Ellerbroek}, {Bouy}  \&
  {Barrado}}{{Bouvier} et~al.}{2013}]{2013Bouvier}
{Bouvier} J.,  {Grankin} K.,  {Ellerbroek} L.~E.,  {Bouy} H.,   {Barrado} D.,
  2013, \mn@doi [\aap] {10.1051/0004-6361/201321389}, \href
  {http://adsabs.harvard.edu/abs/2013A%26A...557A..77B} {557, A77}

\bibitem[\protect\citeauthoryear{{Bozhinova} et~al.,}{{Bozhinova}
  et~al.}{2016}]{2016Bozhinova}
{Bozhinova} I.,  et~al., 2016, \mn@doi [\mnras] {10.1093/mnras/stw2327}, \href
  {http://adsabs.harvard.edu/abs/2016MNRAS.463.4459B} {463, 4459}

\bibitem[\protect\citeauthoryear{{Bryan}, {Maddison}  \& {Liffman}}{{Bryan}
  et~al.}{2019}]{2019Bryan}
{Bryan} G.~R.,  {Maddison} S.~T.,   {Liffman} K.,  2019, \mn@doi [\mnras]
  {10.1093/mnras/stz2401}, \href
  {https://ui.adsabs.harvard.edu/abs/2019MNRAS.489.3879B} {489, 3879}

\bibitem[\protect\citeauthoryear{{Calvet}, {Patino}, {Magris}  \&
  {D'Alessio}}{{Calvet} et~al.}{1991}]{1991Calvet}
{Calvet} N.,  {Patino} A.,  {Magris} G.~C.,   {D'Alessio} P.,  1991, \mn@doi
  [\apj] {10.1086/170618}, \href
  {https://ui.adsabs.harvard.edu/abs/1991ApJ...380..617C} {380, 617}

\bibitem[\protect\citeauthoryear{{Carpenter}, {Hillenbrand}  \&
  {Skrutskie}}{{Carpenter} et~al.}{2001}]{2001Carpenter}
{Carpenter} J.~M.,  {Hillenbrand} L.~A.,   {Skrutskie} M.~F.,  2001, \mn@doi
  [\aj] {10.1086/321086}, \href
  {http://adsabs.harvard.edu/abs/2001AJ....121.3160C} {121, 3160}

\bibitem[\protect\citeauthoryear{{Chambers} et~al.,}{{Chambers}
  et~al.}{2016}]{2016Chambers}
{Chambers} K.~C.,  et~al., 2016, preprint, \href
  {http://adsabs.harvard.edu/abs/2016arXiv161205560C} {} (\mn@eprint {arXiv}
  {1612.05560})

\bibitem[\protect\citeauthoryear{{Cheng}, {Andersen}  \& {Tan}}{{Cheng}
  et~al.}{2020}]{2020Cheng}
{Cheng} Y.,  {Andersen} M.,   {Tan} J.,  2020, \mn@doi [\apj]
  {10.3847/1538-4357/ab93bc}, \href
  {https://ui.adsabs.harvard.edu/abs/2020ApJ...897...51C} {897, 51}

\bibitem[\protect\citeauthoryear{{Cieza} et~al.,}{{Cieza}
  et~al.}{2016}]{2016Cieza}
{Cieza} L.~A.,  et~al., 2016, \mn@doi [\nat] {10.1038/nature18612}, \href
  {http://adsabs.harvard.edu/abs/2016Natur.535..258C} {535, 258}

\bibitem[\protect\citeauthoryear{{Cody} \& {Hillenbrand}}{{Cody} \&
  {Hillenbrand}}{2018}]{2018Cody}
{Cody} A.~M.,  {Hillenbrand} L.~A.,  2018, \mn@doi [\aj]
  {10.3847/1538-3881/aacead}, \href
  {https://ui.adsabs.harvard.edu/abs/2018AJ....156...71C} {156, 71}

\bibitem[\protect\citeauthoryear{{Cody} et~al.,}{{Cody}
  et~al.}{2014}]{2014Cody}
{Cody} A.~M.,  et~al., 2014, \mn@doi [\aj] {10.1088/0004-6256/147/4/82}, \href
  {https://ui.adsabs.harvard.edu/abs/2014AJ....147...82C} {147, 82}

\bibitem[\protect\citeauthoryear{{Connelley} \& {Greene}}{{Connelley} \&
  {Greene}}{2010}]{2010Connelley}
{Connelley} M.~S.,  {Greene} T.~P.,  2010, \mn@doi [\aj]
  {10.1088/0004-6256/140/5/1214}, \href
  {https://ui.adsabs.harvard.edu/abs/2010AJ....140.1214C} {140, 1214}

\bibitem[\protect\citeauthoryear{{Connelley} \& {Reipurth}}{{Connelley} \&
  {Reipurth}}{2018}]{2018Connelley}
{Connelley} M.~S.,  {Reipurth} B.,  2018, \mn@doi [\apj]
  {10.3847/1538-4357/aaba7b}, \href
  {https://ui.adsabs.harvard.edu/abs/2018ApJ...861..145C} {861, 145}

\bibitem[\protect\citeauthoryear{{Connelley} \& {Reipurth}}{{Connelley} \&
  {Reipurth}}{2020}]{2020Connelley}
{Connelley} M.,  {Reipurth} B.,  2020, The Astronomer's Telegram, \href
  {https://ui.adsabs.harvard.edu/abs/2020ATel14035....1C} {14035, 1}

\bibitem[\protect\citeauthoryear{{Contreras Pe{\~n}a} et~al.,}{{Contreras
  Pe{\~n}a} et~al.}{2017a}]{2017Contreras_a}
{Contreras Pe{\~n}a} C.,  et~al., 2017a, \mn@doi [\mnras]
  {10.1093/mnras/stw2801}, \href
  {http://adsabs.harvard.edu/abs/2017MNRAS.465.3011C} {465, 3011}

\bibitem[\protect\citeauthoryear{{Contreras Pe{\~n}a} et~al.,}{{Contreras
  Pe{\~n}a} et~al.}{2017b}]{2017Contreras}
{Contreras Pe{\~n}a} C.,  et~al., 2017b, \mn@doi [\mnras]
  {10.1093/mnras/stw2802}, \href
  {http://adsabs.harvard.edu/abs/2017MNRAS.465.3039C} {465, 3039}

\bibitem[\protect\citeauthoryear{{Contreras Pe{\~n}a}, {Naylor}  \&
  {Morrell}}{{Contreras Pe{\~n}a} et~al.}{2019}]{2019Contreras}
{Contreras Pe{\~n}a} C.,  {Naylor} T.,   {Morrell} S.,  2019, \mn@doi [\mnras]
  {10.1093/mnras/stz1019}, \href
  {https://ui.adsabs.harvard.edu/abs/2019MNRAS.486.4590C} {486, 4590}

\bibitem[\protect\citeauthoryear{{Contreras Pe{\~n}a}, {Johnstone}, {Baek},
  {Herczeg}, {Mairs}, {Scholz}, {Lee}  \& {JCMT Transient Team}}{{Contreras
  Pe{\~n}a} et~al.}{2020}]{2020Contreras}
{Contreras Pe{\~n}a} C.,  {Johnstone} D.,  {Baek} G.,  {Herczeg} G.~J.,
  {Mairs} S.,  {Scholz} A.,  {Lee} J.-E.,   {JCMT Transient Team} 2020, \mn@doi
  [\mnras] {10.1093/mnras/staa1254}, \href
  {https://ui.adsabs.harvard.edu/abs/2020MNRAS.495.3614C} {495, 3614}

\bibitem[\protect\citeauthoryear{{Covey}, {Larson}, {Herczeg}  \&
  {Manara}}{{Covey} et~al.}{2021}]{2021Covey}
{Covey} K.~R.,  {Larson} K.~A.,  {Herczeg} G.~J.,   {Manara} C.~F.,  2021,
  \mn@doi [\aj] {10.3847/1538-3881/abcc73}, \href
  {https://ui.adsabs.harvard.edu/abs/2021AJ....161...61C} {161, 61}

\bibitem[\protect\citeauthoryear{{Cruz-S{\'a}enz de Miera}
  et~al.,}{{Cruz-S{\'a}enz de Miera} et~al.}{2022}]{2022Cruz}
{Cruz-S{\'a}enz de Miera} F.,  et~al., 2022, \mn@doi [\apj]
  {10.3847/1538-4357/ac477f}, \href
  {https://ui.adsabs.harvard.edu/abs/2022ApJ...927..125C} {927, 125}

\bibitem[\protect\citeauthoryear{{Cuello} et~al.,}{{Cuello}
  et~al.}{2019}]{2019Cuello}
{Cuello} N.,  et~al., 2019, \mn@doi [\mnras] {10.1093/mnras/sty3325}, \href
  {https://ui.adsabs.harvard.edu/abs/2019MNRAS.483.4114C} {483, 4114}

\bibitem[\protect\citeauthoryear{{Cushing}, {Vacca}  \& {Rayner}}{{Cushing}
  et~al.}{2004}]{cushing04}
{Cushing} M.~C.,  {Vacca} W.~D.,   {Rayner} J.~T.,  2004, \mn@doi [\pasp]
  {10.1086/382907}, \href
  {https://ui.adsabs.harvard.edu/abs/2004PASP..116..362C} {116, 362}

\bibitem[\protect\citeauthoryear{{Dahm} \& {Hillenbrand}}{{Dahm} \&
  {Hillenbrand}}{2020}]{2020Dahm}
{Dahm} S.~E.,  {Hillenbrand} L.~A.,  2020, \mn@doi [\aj]
  {10.3847/1538-3881/abbfa2}, \href
  {https://ui.adsabs.harvard.edu/abs/2020AJ....160..278D} {160, 278}

\bibitem[\protect\citeauthoryear{{Davis}, {Ray}, {Eisloeffel}  \&
  {Corcoran}}{{Davis} et~al.}{1997}]{1997Davis}
{Davis} C.~J.,  {Ray} T.~P.,  {Eisloeffel} J.,   {Corcoran} D.,  1997, \aap,
  \href {https://ui.adsabs.harvard.edu/abs/1997A&A...324..263D} {324, 263}

\bibitem[\protect\citeauthoryear{{Davis} et~al.,}{{Davis}
  et~al.}{2011}]{2011Davis}
{Davis} C.~J.,  et~al., 2011, \mn@doi [\aap] {10.1051/0004-6361/201015897},
  \href {https://ui.adsabs.harvard.edu/abs/2011A&A...528A...3D} {528, A3}

\bibitem[\protect\citeauthoryear{{Dullemond}, {van den Ancker}, {Acke}  \& {van
  Boekel}}{{Dullemond} et~al.}{2003}]{2003Dullemond}
{Dullemond} C.~P.,  {van den Ancker} M.~E.,  {Acke} B.,   {van Boekel} R.,
  2003, \mn@doi [\apjl] {10.1086/378400}, \href
  {https://ui.adsabs.harvard.edu/abs/2003ApJ...594L..47D} {594, L47}

\bibitem[\protect\citeauthoryear{{Dunham} et~al.,}{{Dunham}
  et~al.}{2015}]{2015Dunham}
{Dunham} M.~M.,  et~al., 2015, \mn@doi [\apjs] {10.1088/0067-0049/220/1/11},
  \href {https://ui.adsabs.harvard.edu/abs/2015ApJS..220...11D} {220, 11}

\bibitem[\protect\citeauthoryear{{Eikenberry} et~al.,}{{Eikenberry}
  et~al.}{2004}]{elkenberry04}
{Eikenberry} S.~S.,  et~al., 2004, in {Moorwood} A. F.~M.,  {Iye} M.,  eds,
  Society of Photo-Optical Instrumentation Engineers (SPIE) Conference Series
  Vol. 5492, Ground-based Instrumentation for Astronomy. pp 1196--1207,
  \mn@doi{10.1117/12.549796}

\bibitem[\protect\citeauthoryear{{Evans} II et~al.,}{{Evans}
  et~al.}{2009}]{2009Evans}
{Evans} II N.~J.,  et~al., 2009, \mn@doi [\apjs] {10.1088/0067-0049/181/2/321},
  \href {http://adsabs.harvard.edu/abs/2009ApJS..181..321E} {181, 321}

\bibitem[\protect\citeauthoryear{{Fiorellino} et~al.,}{{Fiorellino}
  et~al.}{2021}]{2021Fiorellino}
{Fiorellino} E.,  et~al., 2021, \mn@doi [\aap] {10.1051/0004-6361/202039264},
  \href {https://ui.adsabs.harvard.edu/abs/2021A&A...650A..43F} {650, A43}

\bibitem[\protect\citeauthoryear{{Fischer}, {Padgett}, {Stapelfeldt}  \&
  {Sewi{\l}o}}{{Fischer} et~al.}{2016}]{2016Fischer}
{Fischer} W.~J.,  {Padgett} D.~L.,  {Stapelfeldt} K.~L.,   {Sewi{\l}o} M.,
  2016, \mn@doi [\apj] {10.3847/0004-637X/827/2/96}, \href
  {https://ui.adsabs.harvard.edu/abs/2016ApJ...827...96F} {827, 96}

\bibitem[\protect\citeauthoryear{{Fischer}, {Safron}  \& {Megeath}}{{Fischer}
  et~al.}{2019}]{2019Fischer}
{Fischer} W.~J.,  {Safron} E.,   {Megeath} S.~T.,  2019, \mn@doi [\apj]
  {10.3847/1538-4357/ab01dc}, \href
  {http://adsabs.harvard.edu/abs/2019ApJ...872..183F} {872, 183}

\bibitem[\protect\citeauthoryear{{Fischer}, {Hillenbrand}, {Herczeg},
  {Johnstone}, {K{\'o}sp{\'a}l}  \& {Dunham}}{{Fischer}
  et~al.}{2022}]{2022Fischer}
{Fischer} W.~J.,  {Hillenbrand} L.~A.,  {Herczeg} G.~J.,  {Johnstone} D.,
  {K{\'o}sp{\'a}l} {\'A}.,   {Dunham} M.~M.,  2022, arXiv e-prints, \href
  {https://ui.adsabs.harvard.edu/abs/2022arXiv220311257F} {p. arXiv:2203.11257}

\bibitem[\protect\citeauthoryear{{Flaherty}, {Muzerolle}, {Rieke}, {Gutermuth},
  {Balog}, {Herbst}, {Megeath}  \& {Kun}}{{Flaherty}
  et~al.}{2012}]{2012Flaherty}
{Flaherty} K.~M.,  {Muzerolle} J.,  {Rieke} G.,  {Gutermuth} R.,  {Balog} Z.,
  {Herbst} W.,  {Megeath} S.~T.,   {Kun} M.,  2012, \mn@doi [\apj]
  {10.1088/0004-637X/748/1/71}, \href
  {https://ui.adsabs.harvard.edu/abs/2012ApJ...748...71F} {748, 71}

\bibitem[\protect\citeauthoryear{{Francis} et~al.,}{{Francis}
  et~al.}{2022}]{22Francis}
{Francis} L.,  et~al., 2022, \mn@doi [\apj] {10.3847/1538-4357/ac8a9e}, \href
  {https://ui.adsabs.harvard.edu/abs/2022ApJ...937...29F} {937, 29}

\bibitem[\protect\citeauthoryear{{Furlan} et~al.,}{{Furlan}
  et~al.}{2016}]{2016Furlan}
{Furlan} E.,  et~al., 2016, \mn@doi [\apjs] {10.3847/0067-0049/224/1/5}, \href
  {https://ui.adsabs.harvard.edu/abs/2016ApJS..224....5F} {224, 5}

\bibitem[\protect\citeauthoryear{{Galli} et~al.,}{{Galli}
  et~al.}{2021}]{2021Galli}
{Galli} P.~A.~B.,  et~al., 2021, \mn@doi [\aap] {10.1051/0004-6361/202039395},
  \href {https://ui.adsabs.harvard.edu/abs/2021A&A...646A..46G} {646, A46}

\bibitem[\protect\citeauthoryear{{Garc{\'\i}a Soto}, {Ali}, {Newmark},
  {Herbst}, {Windemuth}  \& {Winn}}{{Garc{\'\i}a Soto}
  et~al.}{2020}]{2020Garcia}
{Garc{\'\i}a Soto} A.,  {Ali} A.,  {Newmark} A.,  {Herbst} W.,  {Windemuth} D.,
    {Winn} J.~N.,  2020, \mn@doi [\aj] {10.3847/1538-3881/ab6efd}, \href
  {https://ui.adsabs.harvard.edu/abs/2020AJ....159..135G} {159, 135}

\bibitem[\protect\citeauthoryear{{Ghosh} et~al.,}{{Ghosh}
  et~al.}{2022}]{2022Ghosh}
{Ghosh} A.,  et~al., 2022, \mn@doi [\apj] {10.3847/1538-4357/ac41c2}, \href
  {https://ui.adsabs.harvard.edu/abs/2022ApJ...926...68G} {926, 68}

\bibitem[\protect\citeauthoryear{{G{\'o}mez} \& {Mardones}}{{G{\'o}mez} \&
  {Mardones}}{2003}]{2003Gomez}
{G{\'o}mez} M.,  {Mardones} D.,  2003, \mn@doi [\aj] {10.1086/368391}, \href
  {https://ui.adsabs.harvard.edu/abs/2003AJ....125.2134G} {125, 2134}

\bibitem[\protect\citeauthoryear{{Guo} et~al.,}{{Guo} et~al.}{2018}]{2018Guo}
{Guo} Z.,  et~al., 2018, \mn@doi [\apj] {10.3847/1538-4357/aa9e52}, \href
  {https://ui.adsabs.harvard.edu/abs/2018ApJ...852...56G} {852, 56}

\bibitem[\protect\citeauthoryear{{Guo} et~al.,}{{Guo} et~al.}{2020}]{2020Guo}
{Guo} Z.,  et~al., 2020, \mn@doi [\mnras] {10.1093/mnras/stz3374}, \href
  {https://ui.adsabs.harvard.edu/abs/2020MNRAS.492..294G} {492, 294}

\bibitem[\protect\citeauthoryear{{Guo} et~al.,}{{Guo} et~al.}{2021}]{2021Guo}
{Guo} Z.,  et~al., 2021, \mn@doi [\mnras] {10.1093/mnras/stab882}, \href
  {https://ui.adsabs.harvard.edu/abs/2021MNRAS.504..830G} {504, 830}

\bibitem[\protect\citeauthoryear{{Guo} et~al.,}{{Guo} et~al.}{2022}]{2022Guo}
{Guo} Z.,  et~al., 2022, \mn@doi [\mnras] {10.1093/mnras/stac768}, \href
  {https://ui.adsabs.harvard.edu/abs/2022MNRAS.513.1015G} {513, 1015}

\bibitem[\protect\citeauthoryear{{Gutermuth}, {Megeath}, {Myers}, {Allen},
  {Pipher}  \& {Fazio}}{{Gutermuth} et~al.}{2009}]{2009Gutermuth}
{Gutermuth} R.~A.,  {Megeath} S.~T.,  {Myers} P.~C.,  {Allen} L.~E.,  {Pipher}
  J.~L.,   {Fazio} G.~G.,  2009, \mn@doi [\apjs] {10.1088/0067-0049/184/1/18},
  \href {http://adsabs.harvard.edu/abs/2009ApJS..184...18G} {184, 18}

\bibitem[\protect\citeauthoryear{{Hackstein} et~al.,}{{Hackstein}
  et~al.}{2015}]{2015Hackstein}
{Hackstein} M.,  et~al., 2015, \mn@doi [Astronomische Nachrichten]
  {10.1002/asna.201512195}, \href
  {http://adsabs.harvard.edu/abs/2015AN....336..590H} {336, 590}

\bibitem[\protect\citeauthoryear{{Hambly} et~al.,}{{Hambly}
  et~al.}{2001}]{2001Hambly}
{Hambly} N.~C.,  et~al., 2001, \mn@doi [\mnras]
  {10.1111/j.1365-2966.2001.04660.x}, \href
  {http://adsabs.harvard.edu/abs/2001MNRAS.326.1279H} {326, 1279}

\bibitem[\protect\citeauthoryear{{Hartmann} \& {Kenyon}}{{Hartmann} \&
  {Kenyon}}{1996}]{1996Hartmann}
{Hartmann} L.,  {Kenyon} S.~J.,  1996, \mn@doi [\araa]
  {10.1146/annurev.astro.34.1.207}, \href
  {http://adsabs.harvard.edu/abs/1996ARA%26A..34..207H} {34, 207}

\bibitem[\protect\citeauthoryear{{Herbig}}{{Herbig}}{1977}]{1977Herbig}
{Herbig} G.~H.,  1977, \mn@doi [\apj] {10.1086/155615}, \href
  {https://ui.adsabs.harvard.edu/abs/1977ApJ...217..693H} {217, 693}

\bibitem[\protect\citeauthoryear{{Herbig}}{{Herbig}}{1989}]{1989Herbig}
{Herbig} G.~H.,  1989, in European Southern Observatory Conference and Workshop
  Proceedings. pp 233--246

\bibitem[\protect\citeauthoryear{{Herbig}}{{Herbig}}{2008}]{2008Herbig}
{Herbig} G.~H.,  2008, \mn@doi [\aj] {10.1088/0004-6256/135/2/637}, \href
  {https://ui.adsabs.harvard.edu/abs/2008AJ....135..637H} {135, 637}

\bibitem[\protect\citeauthoryear{{Herbst}, {Herbst}, {Grossman}  \&
  {Weinstein}}{{Herbst} et~al.}{1994}]{1994Herbst}
{Herbst} W.,  {Herbst} D.~K.,  {Grossman} E.~J.,   {Weinstein} D.,  1994,
  \mn@doi [\aj] {10.1086/117204}, \href
  {http://adsabs.harvard.edu/abs/1994AJ....108.1906H} {108, 1906}

\bibitem[\protect\citeauthoryear{{Herczeg} et~al.,}{{Herczeg}
  et~al.}{2017}]{2017Herczeg}
{Herczeg} G.~J.,  et~al., 2017, \mn@doi [\apj] {10.3847/1538-4357/aa8b62},
  \href {https://ui.adsabs.harvard.edu/abs/2017ApJ...849...43H} {849, 43}

\bibitem[\protect\citeauthoryear{{Herter} et~al.,}{{Herter}
  et~al.}{2008}]{herter08}
{Herter} T.~L.,  et~al., 2008, in {McLean} I.~S.,  {Casali} M.~M.,  eds,
  Society of Photo-Optical Instrumentation Engineers (SPIE) Conference Series
  Vol. 7014, Ground-based and Airborne Instrumentation for Astronomy II. p.
  70140X, \mn@doi{10.1117/12.789660}

\bibitem[\protect\citeauthoryear{{Hillenbrand}}{{Hillenbrand}}{2021}]{2021Hillenbrand}
{Hillenbrand} L.~A.,  2021, The Astronomer's Telegram, \href
  {https://ui.adsabs.harvard.edu/abs/2021ATel14590....1H} {14590, 1}

\bibitem[\protect\citeauthoryear{{Hillenbrand} \& {Rodriguez}}{{Hillenbrand} \&
  {Rodriguez}}{2022}]{2022Hillenbrandb}
{Hillenbrand} L.~A.,  {Rodriguez} A.~C.,  2022, \mn@doi [Research Notes of the
  American Astronomical Society] {10.3847/2515-5172/ac4807}, \href
  {https://ui.adsabs.harvard.edu/abs/2022RNAAS...6....6H} {6, 6}

\bibitem[\protect\citeauthoryear{{Hillenbrand} et~al.,}{{Hillenbrand}
  et~al.}{2013}]{2013Hillenbrand}
{Hillenbrand} L.~A.,  et~al., 2013, \mn@doi [\aj] {10.1088/0004-6256/145/3/59},
  \href {http://adsabs.harvard.edu/abs/2013AJ....145...59H} {145, 59}

\bibitem[\protect\citeauthoryear{{Hillenbrand} et~al.,}{{Hillenbrand}
  et~al.}{2018}]{2018Hillenbrand}
{Hillenbrand} L.~A.,  et~al., 2018, The Astrophysical Journal, 869, 146

\bibitem[\protect\citeauthoryear{{Hillenbrand} et~al.,}{{Hillenbrand}
  et~al.}{2021}]{2021Hillenbrand_a}
{Hillenbrand} L.~A.,  et~al., 2021, \mn@doi [\aj] {10.3847/1538-3881/abe406},
  \href {https://ui.adsabs.harvard.edu/abs/2021AJ....161..220H} {161, 220}

\bibitem[\protect\citeauthoryear{{Hillenbrand}, {Isaacson}, {Rodriguez},
  {Connelley}, {Reipurth}, {Kuhn}, {Beck}  \& {Perez}}{{Hillenbrand}
  et~al.}{2022a}]{2022Hillenbrand}
{Hillenbrand} L.~A.,  {Isaacson} H.,  {Rodriguez} A.~C.,  {Connelley} M.,
  {Reipurth} B.,  {Kuhn} M.~A.,  {Beck} T.,   {Perez} D.~R.,  2022a, \mn@doi
  [\aj] {10.3847/1538-3881/ac4752}, \href
  {https://ui.adsabs.harvard.edu/abs/2022AJ....163..115H} {163, 115}

\bibitem[\protect\citeauthoryear{{Hillenbrand}, {Kiker}, {Gee}, {Lester},
  {Braunfeld}, {Rebull}  \& {Kuhn}}{{Hillenbrand}
  et~al.}{2022b}]{2022Hillenbrandc}
{Hillenbrand} L.~A.,  {Kiker} T.~J.,  {Gee} M.,  {Lester} O.,  {Braunfeld}
  N.~L.,  {Rebull} L.~M.,   {Kuhn} M.~A.,  2022b, \mn@doi [\aj]
  {10.3847/1538-3881/ac62d8}, \href
  {https://ui.adsabs.harvard.edu/abs/2022AJ....163..263H} {163, 263}

\bibitem[\protect\citeauthoryear{{Hodapp} \& {Chini}}{{Hodapp} \&
  {Chini}}{2015}]{2015Hodapp}
{Hodapp} K.~W.,  {Chini} R.,  2015, \mn@doi [\apj]
  {10.1088/0004-637X/813/2/107}, \href
  {https://ui.adsabs.harvard.edu/abs/2015ApJ...813..107H} {813, 107}

\bibitem[\protect\citeauthoryear{{Hodapp}, {Hora}, {Rayner}, {Pickles}  \&
  {Ladd}}{{Hodapp} et~al.}{1996}]{1996Hodapp}
{Hodapp} K.-W.,  {Hora} J.~L.,  {Rayner} J.~T.,  {Pickles} A.~J.,   {Ladd}
  E.~F.,  1996, \mn@doi [\apj] {10.1086/177742}, \href
  {https://ui.adsabs.harvard.edu/abs/1996ApJ...468..861H} {468, 861}

\bibitem[\protect\citeauthoryear{{Hodapp}, {Chini}, {Watermann}  \&
  {Lemke}}{{Hodapp} et~al.}{2012}]{2012Hodapp}
{Hodapp} K.~W.,  {Chini} R.,  {Watermann} R.,   {Lemke} R.,  2012, \mn@doi
  [\apj] {10.1088/0004-637X/744/1/56}, \href
  {https://ui.adsabs.harvard.edu/abs/2012ApJ...744...56H} {744, 56}

\bibitem[\protect\citeauthoryear{{Hodapp} et~al.,}{{Hodapp}
  et~al.}{2019}]{2019Hodapp}
{Hodapp} K.~W.,  et~al., 2019, \mn@doi [\aj] {10.3847/1538-3881/ab471a}, \href
  {https://ui.adsabs.harvard.edu/abs/2019AJ....158..241H} {158, 241}

\bibitem[\protect\citeauthoryear{{Hodapp} et~al.,}{{Hodapp}
  et~al.}{2020}]{2020Hodapp}
{Hodapp} K.~W.,  et~al., 2020, \mn@doi [\aj] {10.3847/1538-3881/abad96}, \href
  {https://ui.adsabs.harvard.edu/abs/2020AJ....160..164H} {160, 164}

\bibitem[\protect\citeauthoryear{{Hodgkin} et~al.,}{{Hodgkin}
  et~al.}{2021}]{2021Hodgkin}
{Hodgkin} S.~T.,  et~al., 2021, \mn@doi [\aap] {10.1051/0004-6361/202140735},
  \href {https://ui.adsabs.harvard.edu/abs/2021A&A...652A..76H} {652, A76}

\bibitem[\protect\citeauthoryear{{Jayasinghe} et~al.,}{{Jayasinghe}
  et~al.}{2018}]{2018Jayasinghe}
{Jayasinghe} T.,  et~al., 2018, \mn@doi [\mnras] {10.1093/mnras/sty838}, \href
  {https://ui.adsabs.harvard.edu/abs/2018MNRAS.477.3145J} {477, 3145}

\bibitem[\protect\citeauthoryear{{Johnstone}, {Hendricks}, {Herczeg}  \&
  {Bruderer}}{{Johnstone} et~al.}{2013}]{Johnstone13}
{Johnstone} D.,  {Hendricks} B.,  {Herczeg} G.~J.,   {Bruderer} S.,  2013,
  \mn@doi [\apj] {10.1088/0004-637X/765/2/133}, \href
  {https://ui.adsabs.harvard.edu/abs/2013ApJ...765..133J} {765, 133}

\bibitem[\protect\citeauthoryear{{Johnstone} et~al.,}{{Johnstone}
  et~al.}{2022}]{2022Johnstone}
{Johnstone} D.,  et~al., 2022, \mn@doi [\apj] {10.3847/1538-4357/ac8a48}, \href
  {https://ui.adsabs.harvard.edu/abs/2022ApJ...937....6J} {937, 6}

\bibitem[\protect\citeauthoryear{{Kadam}, {Vorobyov}, {Reg{\'a}ly},
  {K{\'o}sp{\'a}l}  \& {{\'A}brah{\'a}m}}{{Kadam} et~al.}{2020}]{2020Kadam}
{Kadam} K.,  {Vorobyov} E.,  {Reg{\'a}ly} Z.,  {K{\'o}sp{\'a}l} {\'A}.,
  {{\'A}brah{\'a}m} P.,  2020, \mn@doi [\apj] {10.3847/1538-4357/ab8bd8}, \href
  {https://ui.adsabs.harvard.edu/abs/2020ApJ...895...41K} {895, 41}

\bibitem[\protect\citeauthoryear{{Kim}, {Kawamura}, {Yonekura}  \&
  {Fukui}}{{Kim} et~al.}{2004}]{2004Kim}
{Kim} B.~G.,  {Kawamura} A.,  {Yonekura} Y.,   {Fukui} Y.,  2004, \mn@doi
  [\pasj] {10.1093/pasj/56.2.313}, \href
  {https://ui.adsabs.harvard.edu/abs/2004PASJ...56..313K} {56, 313}

\bibitem[\protect\citeauthoryear{{K{\'o}sp{\'a}l}, {{\'A}brah{\'a}m}, {Prusti},
  {Acosta-Pulido}, {Hony}, {Mo{\'o}r}  \& {Siebenmorgen}}{{K{\'o}sp{\'a}l}
  et~al.}{2007}]{2007Kospal}
{K{\'o}sp{\'a}l} {\'A}.,  {{\'A}brah{\'a}m} P.,  {Prusti} T.,  {Acosta-Pulido}
  J.,  {Hony} S.,  {Mo{\'o}r} A.,   {Siebenmorgen} R.,  2007, \mn@doi [\aap]
  {10.1051/0004-6361:20066108}, \href
  {https://ui.adsabs.harvard.edu/abs/2007A&A...470..211K} {470, 211}

\bibitem[\protect\citeauthoryear{{Kryukova}, {Megeath}, {Gutermuth}, {Pipher},
  {Allen}, {Allen}, {Myers}  \& {Muzerolle}}{{Kryukova}
  et~al.}{2012}]{2012Kryukova}
{Kryukova} E.,  {Megeath} S.~T.,  {Gutermuth} R.~A.,  {Pipher} J.,  {Allen}
  T.~S.,  {Allen} L.~E.,  {Myers} P.~C.,   {Muzerolle} J.,  2012, \mn@doi [\aj]
  {10.1088/0004-6256/144/2/31}, \href
  {https://ui.adsabs.harvard.edu/abs/2012AJ....144...31K} {144, 31}

\bibitem[\protect\citeauthoryear{{Kun}, {{\'A}brah{\'a}m}, {Acosta Pulido},
  {Mo{\'o}r}  \& {Prusti}}{{Kun} et~al.}{2019}]{2019Kun}
{Kun} M.,  {{\'A}brah{\'a}m} P.,  {Acosta Pulido} J.~A.,  {Mo{\'o}r} A.,
  {Prusti} T.,  2019, \mn@doi [\mnras] {10.1093/mnras/sty3425}, \href
  {https://ui.adsabs.harvard.edu/abs/2019MNRAS.483.4424K} {483, 4424}

\bibitem[\protect\citeauthoryear{{Kunitomo}, {Guillot}, {Takeuchi}  \&
  {Ida}}{{Kunitomo} et~al.}{2017}]{2017Kunimoto}
{Kunitomo} M.,  {Guillot} T.,  {Takeuchi} T.,   {Ida} S.,  2017, \mn@doi [\aap]
  {10.1051/0004-6361/201628260}, \href
  {https://ui.adsabs.harvard.edu/abs/2017A&A...599A..49K} {599, A49}

\bibitem[\protect\citeauthoryear{{Lada}, {Alves}  \& {Lada}}{{Lada}
  et~al.}{1996}]{1996Lada}
{Lada} C.~J.,  {Alves} J.,   {Lada} E.~A.,  1996, \mn@doi [\aj]
  {10.1086/117933}, \href
  {https://ui.adsabs.harvard.edu/abs/1996AJ....111.1964L} {111, 1964}

\bibitem[\protect\citeauthoryear{{Lee} et~al.,}{{Lee} et~al.}{2019}]{2019Lee}
{Lee} J.-E.,  et~al., 2019, \mn@doi [Nature Astronomy]
  {10.1038/s41550-018-0680-0}, \href
  {https://ui.adsabs.harvard.edu/abs/2019NatAs...3..314L} {3, 314}

\bibitem[\protect\citeauthoryear{{Lee} et~al.,}{{Lee} et~al.}{2020}]{2020YHLee}
{Lee} Y.-H.,  et~al., 2020, \mn@doi [\apj] {10.3847/1538-4357/abb6fe}, \href
  {https://ui.adsabs.harvard.edu/abs/2020ApJ...903....5L} {903, 5}

\bibitem[\protect\citeauthoryear{{Lee} et~al.,}{{Lee} et~al.}{2021}]{2021Lee}
{Lee} Y.-H.,  et~al., 2021, \mn@doi [\apj] {10.3847/1538-4357/ac1679}, \href
  {https://ui.adsabs.harvard.edu/abs/2021ApJ...920..119L} {920, 119}

\bibitem[\protect\citeauthoryear{{Liu} et~al.,}{{Liu} et~al.}{2022}]{2022Liu}
{Liu} H.,  et~al., 2022, \mn@doi [\apj] {10.3847/1538-4357/ac84d2}, \href
  {https://ui.adsabs.harvard.edu/abs/2022ApJ...936..152L} {936, 152}

\bibitem[\protect\citeauthoryear{{Lodato} \& {Clarke}}{{Lodato} \&
  {Clarke}}{2004}]{2004Lodato}
{Lodato} G.,  {Clarke} C.~J.,  2004, \mn@doi [\mnras]
  {10.1111/j.1365-2966.2004.08112.x}, \href
  {http://adsabs.harvard.edu/abs/2004MNRAS.353..841L} {353, 841}

\bibitem[\protect\citeauthoryear{{Lomb}}{{Lomb}}{1976}]{1976Lomb}
{Lomb} N.~R.,  1976, \mn@doi [\apss] {10.1007/BF00648343}, \href
  {https://ui.adsabs.harvard.edu/abs/1976Ap&SS..39..447L} {39, 447}

\bibitem[\protect\citeauthoryear{{Long} et~al.,}{{Long}
  et~al.}{2017}]{2017Long}
{Long} F.,  et~al., 2017, \mn@doi [\apj] {10.3847/1538-4357/aa78fc}, \href
  {https://ui.adsabs.harvard.edu/abs/2017ApJ...844...99L} {844, 99}

\bibitem[\protect\citeauthoryear{{Lorenzetti}, {Larionov}, {Giannini},
  {Arkharov}, {Antoniucci}, {Nisini}  \& {Di Paola}}{{Lorenzetti}
  et~al.}{2009}]{2009Lorenzetti}
{Lorenzetti} D.,  {Larionov} V.~M.,  {Giannini} T.,  {Arkharov} A.~A.,
  {Antoniucci} S.,  {Nisini} B.,   {Di Paola} A.,  2009, \mn@doi [\apj]
  {10.1088/0004-637X/693/2/1056}, \href
  {https://ui.adsabs.harvard.edu/abs/2009ApJ...693.1056L} {693, 1056}

\bibitem[\protect\citeauthoryear{{Lorenzetti} et~al.,}{{Lorenzetti}
  et~al.}{2012}]{2012Lorenzetti}
{Lorenzetti} D.,  et~al., 2012, \mn@doi [\apj] {10.1088/0004-637X/749/2/188},
  \href {https://ui.adsabs.harvard.edu/abs/2012ApJ...749..188L} {749, 188}

\bibitem[\protect\citeauthoryear{{Lucas} et~al.,}{{Lucas}
  et~al.}{2008}]{c2008Lucas}
{Lucas} P.~W.,  et~al., 2008, \mn@doi [\mnras]
  {10.1111/j.1365-2966.2008.13924.x}, \href
  {https://ui.adsabs.harvard.edu/\#abs/2008MNRAS.391..136L} {391, 136}

\bibitem[\protect\citeauthoryear{{Lucas} et~al.,}{{Lucas}
  et~al.}{2020}]{2020Lucas}
{Lucas} P.~W.,  et~al., 2020, \mn@doi [\mnras] {10.1093/mnras/staa2915}, \href
  {https://ui.adsabs.harvard.edu/abs/2020MNRAS.499.1805L} {499, 1805}

\bibitem[\protect\citeauthoryear{{Maddalena}, {Morris}, {Moskowitz}  \&
  {Thaddeus}}{{Maddalena} et~al.}{1986}]{1986Maddalena}
{Maddalena} R.~J.,  {Morris} M.,  {Moskowitz} J.,   {Thaddeus} P.,  1986,
  \mn@doi [\apj] {10.1086/164083}, \href
  {https://ui.adsabs.harvard.edu/abs/1986ApJ...303..375M} {303, 375}

\bibitem[\protect\citeauthoryear{{Magakian}, {Movsessian}  \&
  {Nikogossian}}{{Magakian} et~al.}{2008}]{2008Magakian}
{Magakian} T.~Y.,  {Movsessian} T.~A.,   {Nikogossian} E.~G.,  2008, \mn@doi
  [Astrophysics] {10.1007/s10511-008-0002-9}, \href
  {https://ui.adsabs.harvard.edu/abs/2008Ap.....51....7M} {51, 7}

\bibitem[\protect\citeauthoryear{{Mainzer} et~al.,}{{Mainzer}
  et~al.}{2011}]{2011Mainzer}
{Mainzer} A.,  et~al., 2011, \mn@doi [\apj] {10.1088/0004-637X/743/2/156},
  \href {https://ui.adsabs.harvard.edu/abs/2011ApJ...743..156M} {743, 156}

\bibitem[\protect\citeauthoryear{{Mainzer} et~al.,}{{Mainzer}
  et~al.}{2014}]{2014Mainzer}
{Mainzer} A.,  et~al., 2014, \mn@doi [\apj] {10.1088/0004-637X/792/1/30}, \href
  {https://ui.adsabs.harvard.edu/abs/2014ApJ...792...30M} {792, 30}

\bibitem[\protect\citeauthoryear{Mairs et~al.,}{Mairs
  et~al.}{2017a}]{2017Mairs}
Mairs S.,  et~al., 2017a, \mn@doi [The Astrophysical Journal]
  {10.3847/1538-4357/aa7844}, 843, 55

\bibitem[\protect\citeauthoryear{{Mairs} et~al.,}{{Mairs}
  et~al.}{2017b}]{mairs17b}
{Mairs} S.,  et~al., 2017b, \mn@doi [\apj] {10.3847/1538-4357/aa9225}, \href
  {https://ui.adsabs.harvard.edu/abs/2017ApJ...849..107M} {849, 107}

\bibitem[\protect\citeauthoryear{{Megeath} et~al.,}{{Megeath}
  et~al.}{2012}]{2012Megeath}
{Megeath} S.~T.,  et~al., 2012, \mn@doi [\aj] {10.1088/0004-6256/144/6/192},
  \href {http://adsabs.harvard.edu/abs/2012AJ....144..192M} {144, 192}

\bibitem[\protect\citeauthoryear{{Moore} et~al.,}{{Moore}
  et~al.}{2016}]{2016Moore}
{Moore} A.~M.,  et~al., 2016, in {Hall} H.~J.,  {Gilmozzi} R.,   {Marshall}
  H.~K.,  eds,  Society of Photo-Optical Instrumentation Engineers (SPIE)
  Conference Series Vol. 9906, Ground-based and Airborne Telescopes VI. p.
  99062C (\mn@eprint {arXiv} {1608.04510}), \mn@doi{10.1117/12.2233694}

\bibitem[\protect\citeauthoryear{{Morales-Calder{\'o}n}
  et~al.,}{{Morales-Calder{\'o}n} et~al.}{2011}]{c2011Morales}
{Morales-Calder{\'o}n} M.,  et~al., 2011, \mn@doi [\apj]
  {10.1088/0004-637X/733/1/50}, \href
  {https://ui.adsabs.harvard.edu/\#abs/2011ApJ...733...50M} {733, 50}

\bibitem[\protect\citeauthoryear{{Nikoghosyan}, {Azatyan}  \&
  {Khachatryan}}{{Nikoghosyan} et~al.}{2017}]{2017Niko}
{Nikoghosyan} E.~H.,  {Azatyan} N.~M.,   {Khachatryan} K.~G.,  2017, \mn@doi
  [\aap] {10.1051/0004-6361/201629415}, \href
  {https://ui.adsabs.harvard.edu/abs/2017A&A...603A..26N} {603, A26}

\bibitem[\protect\citeauthoryear{{Park} et~al.,}{{Park}
  et~al.}{2021}]{2021Park}
{Park} W.,  et~al., 2021, \mn@doi [\apj] {10.3847/1538-4357/ac1745}, \href
  {https://ui.adsabs.harvard.edu/abs/2021ApJ...920..132P} {920, 132}

\bibitem[\protect\citeauthoryear{{Park} et~al.,}{{Park}
  et~al.}{2022}]{2022Park}
{Park} S.,  et~al., 2022, arXiv e-prints, \href
  {https://ui.adsabs.harvard.edu/abs/2022arXiv221102137P} {p. arXiv:2211.02137}

\bibitem[\protect\citeauthoryear{{Persi}, {Tapia}, {G{\`o}mez}, {Whitney},
  {Marenzi}  \& {Roth}}{{Persi} et~al.}{2007}]{2007Persi}
{Persi} P.,  {Tapia} M.,  {G{\`o}mez} M.,  {Whitney} B.~A.,  {Marenzi} A.~R.,
  {Roth} M.,  2007, \mn@doi [\aj] {10.1086/511771}, \href
  {https://ui.adsabs.harvard.edu/abs/2007AJ....133.1690P} {133, 1690}

\bibitem[\protect\citeauthoryear{{Rayner}, {Toomey}, {Onaka}, {Denault},
  {Stahlberger}, {Vacca}, {Cushing}  \& {Wang}}{{Rayner}
  et~al.}{2003}]{rayner03}
{Rayner} J.~T.,  {Toomey} D.~W.,  {Onaka} P.~M.,  {Denault} A.~J.,
  {Stahlberger} W.~E.,  {Vacca} W.~D.,  {Cushing} M.~C.,   {Wang} S.,  2003,
  \mn@doi [\pasp] {10.1086/367745}, \href
  {https://ui.adsabs.harvard.edu/abs/2003PASP..115..362R} {115, 362}

\bibitem[\protect\citeauthoryear{{Reipurth} \& {Aspin}}{{Reipurth} \&
  {Aspin}}{2010}]{2010Reipurth_b}
{Reipurth} B.,  {Aspin} C.,  2010, in {Harutyunian} H.~A.,  {Mickaelian} A.~M.,
    {Terzian} Y.,  eds, Evolution of Cosmic Objects through their Physical
  Activity. pp 19--38

\bibitem[\protect\citeauthoryear{{Reipurth}, {Herbig}  \& {Aspin}}{{Reipurth}
  et~al.}{2010}]{2010Reipurth}
{Reipurth} B.,  {Herbig} G.,   {Aspin} C.,  2010, \mn@doi [\aj]
  {10.1088/0004-6256/139/4/1668}, \href
  {http://adsabs.harvard.edu/abs/2010AJ....139.1668R} {139, 1668}

\bibitem[\protect\citeauthoryear{{Safron} et~al.,}{{Safron}
  et~al.}{2015}]{2015Safron}
{Safron} E.~J.,  et~al., 2015, \mn@doi [\apjl] {10.1088/2041-8205/800/1/L5},
  \href {https://ui.adsabs.harvard.edu/abs/2015ApJ...800L...5S} {800, L5}

\bibitem[\protect\citeauthoryear{{Saito} et~al.,}{{Saito}
  et~al.}{2012}]{2012Saito}
{Saito} R.~K.,  et~al., 2012, \mn@doi [\aap] {10.1051/0004-6361/201118407},
  \href {http://adsabs.harvard.edu/abs/2012A%26A...537A.107S} {537, A107}

\bibitem[\protect\citeauthoryear{{Scargle}}{{Scargle}}{1989}]{1989Scargle}
{Scargle} J.~D.,  1989, \mn@doi [\apj] {10.1086/167757}, \href
  {https://ui.adsabs.harvard.edu/abs/1989ApJ...343..874S} {343, 874}

\bibitem[\protect\citeauthoryear{{Scholz}, {Froebrich}  \& {Wood}}{{Scholz}
  et~al.}{2013}]{2013Scholz}
{Scholz} A.,  {Froebrich} D.,   {Wood} K.,  2013, \mn@doi [\mnras]
  {10.1093/mnras/stt091}, \href
  {http://adsabs.harvard.edu/abs/2013MNRAS.430.2910S} {430, 2910}

\bibitem[\protect\citeauthoryear{{Sergison}, {Naylor}, {Littlefair}, {Bell}  \&
  {Williams}}{{Sergison} et~al.}{2020}]{2020Sergison}
{Sergison} D.~J.,  {Naylor} T.,  {Littlefair} S.~P.,  {Bell} C. P.~M.,
  {Williams} C.~D.~H.,  2020, \mn@doi [\mnras] {10.1093/mnras/stz3398}, \href
  {https://ui.adsabs.harvard.edu/abs/2020MNRAS.491.5035S} {491, 5035}

\bibitem[\protect\citeauthoryear{{Sicilia-Aguilar} et~al.,}{{Sicilia-Aguilar}
  et~al.}{2017}]{2017Sicilia}
{Sicilia-Aguilar} A.,  et~al., 2017, \mn@doi [\aap]
  {10.1051/0004-6361/201731263}, \href
  {https://ui.adsabs.harvard.edu/abs/2017A&A...607A.127S} {607, A127}

\bibitem[\protect\citeauthoryear{{Simon}, {Andrews}, {Rayner}  \&
  {Drake}}{{Simon} et~al.}{2004}]{2004Simon}
{Simon} T.,  {Andrews} S.~M.,  {Rayner} J.~T.,   {Drake} S.~A.,  2004, \mn@doi
  [\apj] {10.1086/422187}, \href
  {https://ui.adsabs.harvard.edu/abs/2004ApJ...611..940S} {611, 940}

\bibitem[\protect\citeauthoryear{{Stamatellos}, {Whitworth}  \&
  {Hubber}}{{Stamatellos} et~al.}{2012}]{2012Stamatellos}
{Stamatellos} D.,  {Whitworth} A.~P.,   {Hubber} D.~A.,  2012, \mn@doi [\mnras]
  {10.1111/j.1365-2966.2012.22038.x}, \href
  {https://ui.adsabs.harvard.edu/abs/2012MNRAS.427.1182S} {427, 1182}

\bibitem[\protect\citeauthoryear{{Stauffer} et~al.,}{{Stauffer}
  et~al.}{2014}]{2014Stauffer}
{Stauffer} J.,  et~al., 2014, \mn@doi [\aj] {10.1088/0004-6256/147/4/83}, \href
  {https://ui.adsabs.harvard.edu/abs/2014AJ....147...83S} {147, 83}

\bibitem[\protect\citeauthoryear{Stauffer et~al.,}{Stauffer
  et~al.}{2015}]{2015Stauffer}
Stauffer J.,  et~al., 2015, \mn@doi [The Astronomical Journal]
  {10.1088/0004-6256/149/4/130}, 149, 130

\bibitem[\protect\citeauthoryear{{Szegedi-Elek} et~al.,}{{Szegedi-Elek}
  et~al.}{2020}]{2020Szegedi-Elek}
{Szegedi-Elek} E.,  et~al., 2020, \mn@doi [\apj] {10.3847/1538-4357/aba129},
  \href {https://ui.adsabs.harvard.edu/abs/2020ApJ...899..130S} {899, 130}

\bibitem[\protect\citeauthoryear{{Szil{\'a}gyi}, {Kun}  \&
  {{\'A}brah{\'a}m}}{{Szil{\'a}gyi} et~al.}{2021}]{2021Szi}
{Szil{\'a}gyi} M.,  {Kun} M.,   {{\'A}brah{\'a}m} P.,  2021, \mn@doi [\mnras]
  {10.1093/mnras/stab1496}, \href
  {https://ui.adsabs.harvard.edu/abs/2021MNRAS.505.5164S} {505, 5164}

\bibitem[\protect\citeauthoryear{{Tapia}, {Roth}  \& {Persi}}{{Tapia}
  et~al.}{2015}]{2015Tapia}
{Tapia} M.,  {Roth} M.,   {Persi} P.,  2015, \mn@doi [\mnras]
  {10.1093/mnras/stu2362}, \href
  {https://ui.adsabs.harvard.edu/abs/2015MNRAS.446.4088T} {446, 4088}

\bibitem[\protect\citeauthoryear{{Tobin} et~al.,}{{Tobin}
  et~al.}{2020}]{2020Tobin}
{Tobin} J.~J.,  et~al., 2020, \mn@doi [\apj] {10.3847/1538-4357/ab6f64}, \href
  {https://ui.adsabs.harvard.edu/abs/2020ApJ...890..130T} {890, 130}

\bibitem[\protect\citeauthoryear{{Venuti} et~al.,}{{Venuti}
  et~al.}{2017}]{2017Venuti}
{Venuti} L.,  et~al., 2017, \mn@doi [\aap] {10.1051/0004-6361/201629537}, \href
  {https://ui.adsabs.harvard.edu/abs/2017A&A...599A..23V} {599, A23}

\bibitem[\protect\citeauthoryear{{Vorobyov} \& {Basu}}{{Vorobyov} \&
  {Basu}}{2015}]{2015Vorobyov}
{Vorobyov} E.~I.,  {Basu} S.,  2015, \mn@doi [\apj]
  {10.1088/0004-637X/805/2/115}, \href
  {http://adsabs.harvard.edu/abs/2015ApJ...805..115V} {805, 115}

\bibitem[\protect\citeauthoryear{{Wang} \& {Chen}}{{Wang} \&
  {Chen}}{2019}]{2019Wang}
{Wang} S.,  {Chen} X.,  2019, \mn@doi [\apj] {10.3847/1538-4357/ab1c61}, \href
  {https://ui.adsabs.harvard.edu/abs/2019ApJ...877..116W} {877, 116}

\bibitem[\protect\citeauthoryear{{Wang}, {Lai}, {Clemens}, {Koch}, {Eswaraiah},
  {Chen}  \& {Pandey}}{{Wang} et~al.}{2020}]{2020Wang}
{Wang} J.-W.,  {Lai} S.-P.,  {Clemens} D.~P.,  {Koch} P.~M.,  {Eswaraiah} C.,
  {Chen} W.-P.,   {Pandey} A.~K.,  2020, \mn@doi [\apj]
  {10.3847/1538-4357/ab5c1c}, \href
  {https://ui.adsabs.harvard.edu/abs/2020ApJ...888...13W} {888, 13}

\bibitem[\protect\citeauthoryear{{Ward-Thompson} et~al.,}{{Ward-Thompson}
  et~al.}{2007}]{2007Ward}
{Ward-Thompson} D.,  et~al., 2007, \mn@doi [\pasp] {10.1086/521277}, \href
  {https://ui.adsabs.harvard.edu/abs/2007PASP..119..855W} {119, 855}

\bibitem[\protect\citeauthoryear{{Wright} et~al.,}{{Wright}
  et~al.}{2010}]{2010Wright}
{Wright} E.~L.,  et~al., 2010, \mn@doi [AJ] {10.1088/0004-6256/140/6/1868},
  \href {http://adsabs.harvard.edu/abs/2010AJ....140.1868W} {140, 1868}

\bibitem[\protect\citeauthoryear{{Yoo} et~al.,}{{Yoo} et~al.}{2017}]{2017Yoo}
{Yoo} H.,  et~al., 2017, \mn@doi [\apj] {10.3847/1538-4357/aa8c0a}, \href
  {https://ui.adsabs.harvard.edu/abs/2017ApJ...849...69Y} {849, 69}

\bibitem[\protect\citeauthoryear{{Yoon} et~al.,}{{Yoon}
  et~al.}{2022}]{2022Yoon}
{Yoon} S.-Y.,  et~al., 2022, \mn@doi [\apj] {10.3847/1538-4357/ac5632}, \href
  {https://ui.adsabs.harvard.edu/abs/2022ApJ...929...60Y} {929, 60}

\bibitem[\protect\citeauthoryear{{Zakri} et~al.,}{{Zakri}
  et~al.}{2022}]{2022Zakri}
{Zakri} W.,  et~al., 2022, \mn@doi [\apjl] {10.3847/2041-8213/ac46ae}, \href
  {https://ui.adsabs.harvard.edu/abs/2022ApJ...924L..23Z} {924, L23}

\bibitem[\protect\citeauthoryear{{Zhu}, {Hartmann}  \& {Gammie}}{{Zhu}
  et~al.}{2009}]{2009Zhu}
{Zhu} Z.,  {Hartmann} L.,   {Gammie} C.,  2009, \mn@doi [\apj]
  {10.1088/0004-637X/694/2/1045}, \href
  {http://adsabs.harvard.edu/abs/2009ApJ...694.1045Z} {694, 1045}

\bibitem[\protect\citeauthoryear{{Zhu} et~al.,}{{Zhu} et~al.}{2022}]{2022Wei}
{Zhu} W.,  et~al., 2022, \mn@doi [\apjl] {10.3847/2041-8213/ac7b2d}, \href
  {https://ui.adsabs.harvard.edu/abs/2022ApJ...933L..21Z} {933, L21}

\bibitem[\protect\citeauthoryear{{Zucker}, {Schlafly}, {Speagle}, {Green},
  {Portillo}, {Finkbeiner}  \& {Goodman}}{{Zucker} et~al.}{2018}]{2018Zucker}
{Zucker} C.,  {Schlafly} E.~F.,  {Speagle} J.~S.,  {Green} G.~M.,  {Portillo}
  S. K.~N.,  {Finkbeiner} D.~P.,   {Goodman} A.~A.,  2018, \mn@doi [\apj]
  {10.3847/1538-4357/aae97c}, \href
  {https://ui.adsabs.harvard.edu/abs/2018ApJ...869...83Z} {869, 83}

\makeatother
\end{thebibliography}




\appendix

\section{Individual sources}\label{app:a}

\begin{figure*}
	\resizebox{\columnwidth}{!}{\includegraphics[angle=0]{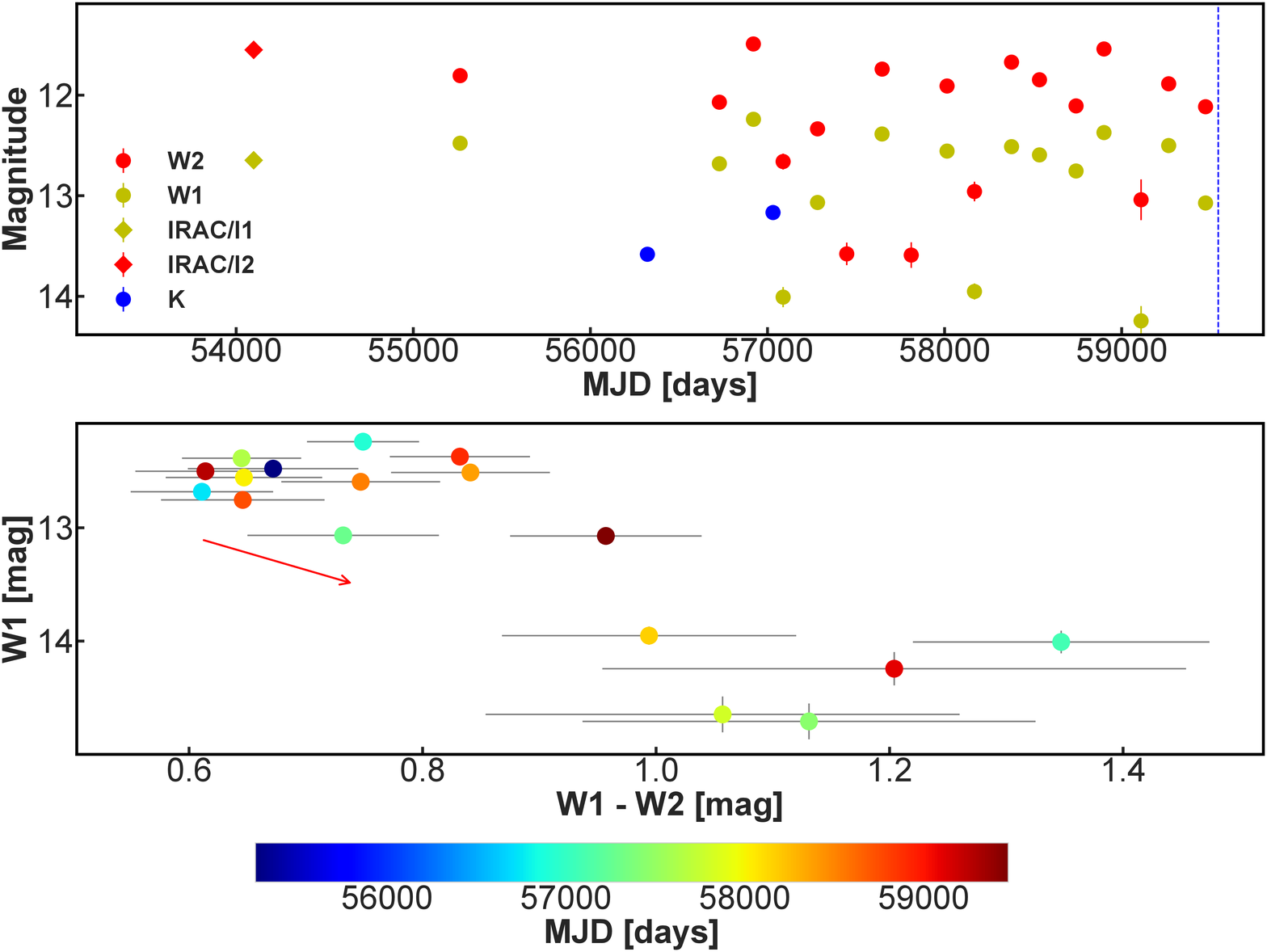}}
	\resizebox{\columnwidth}{!}{\includegraphics[angle=0]{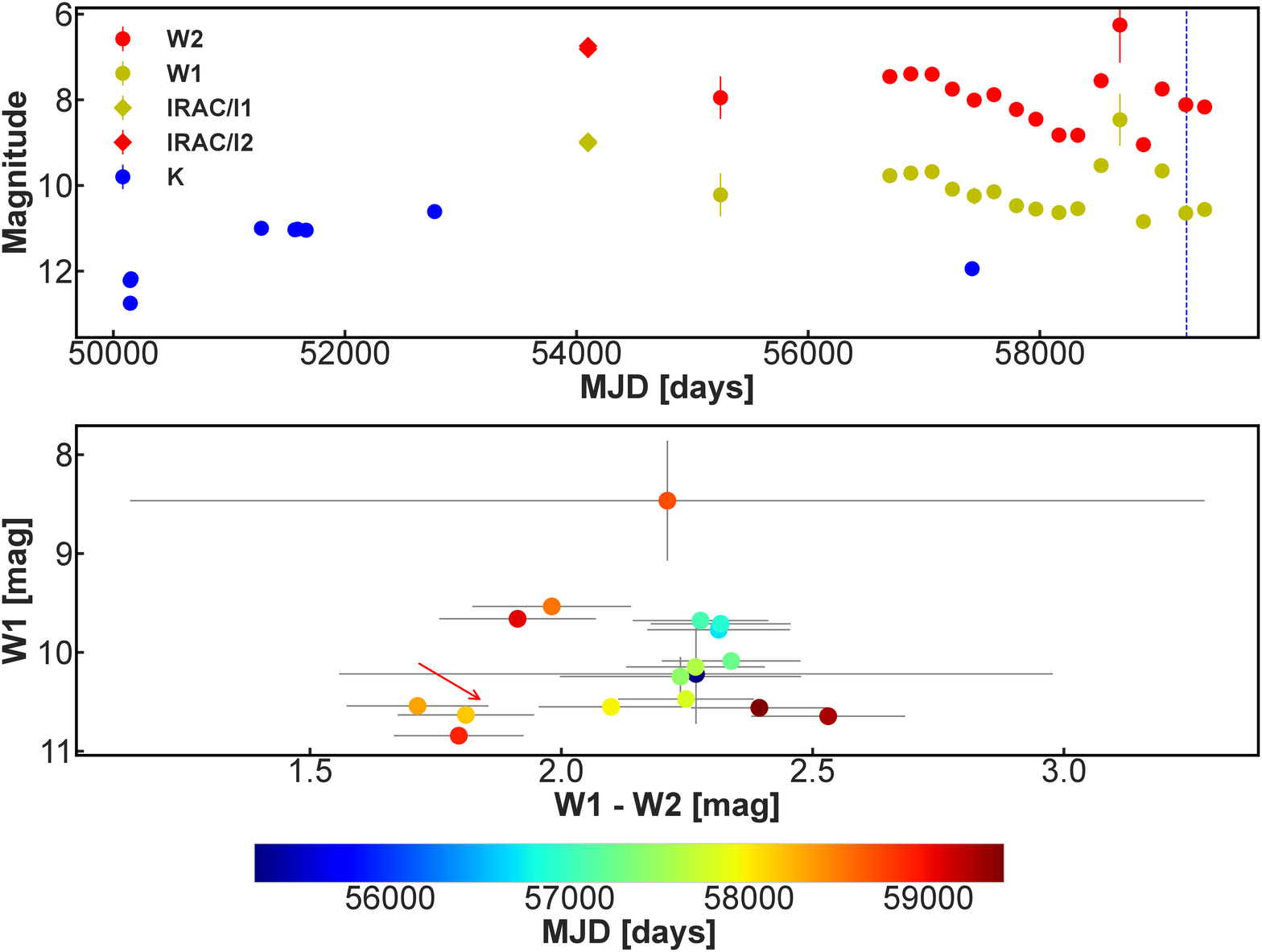}}\\
	\resizebox{\columnwidth}{!}{\includegraphics[angle=0]{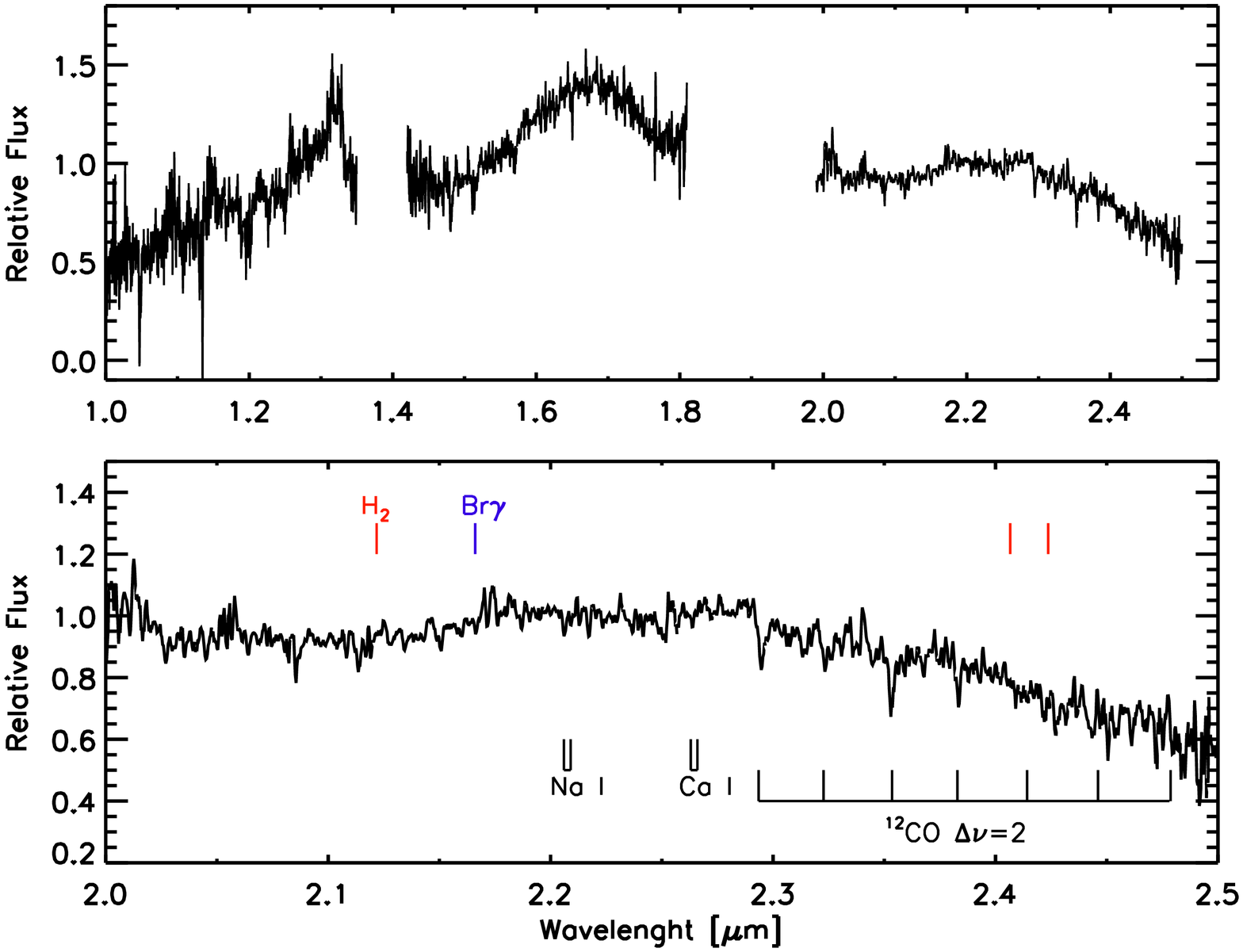}}
	\resizebox{\columnwidth}{!}{\includegraphics[angle=0]{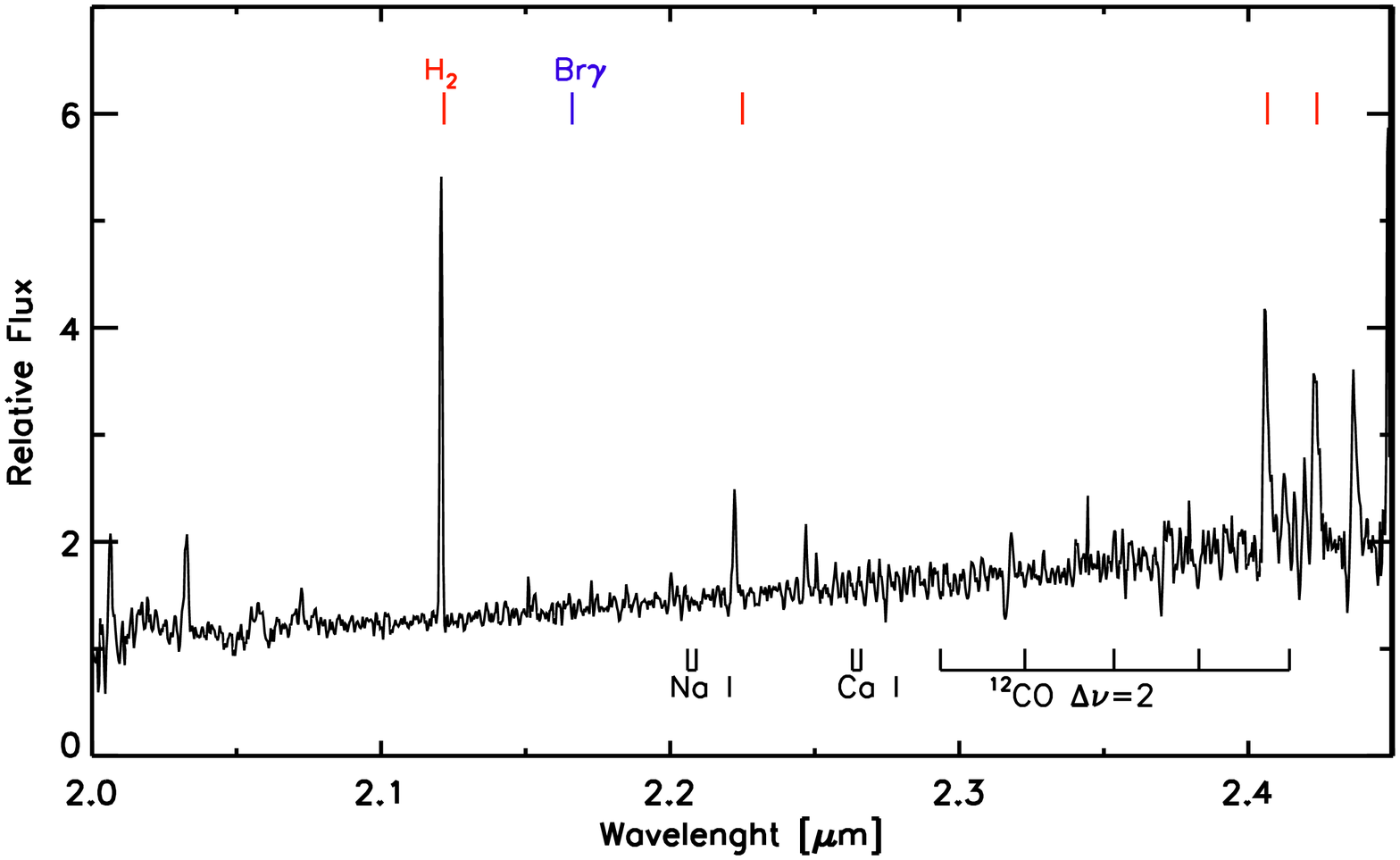}}
	 \caption{(top) K-band (blue), 3.4 $\mu$m (yellow) and 4.6 $\mu$m (red) light curve of HOPS 154 (left) and GM Cha (right). The date of spectroscopic observations is marked by a dashed blue line. For GM Cha we also mark the date of spectroscopic observations from \citet{2003Gomez}. (middle) $W1$ vs $W1-W2$ colour-magnitude diagram for HOPS 154 (left) and GM Cha (right). The plots only include data obtained by the {\it WISE} telescope. The red arrow marks the reddening line for A$_{V}=20$~mag, using the extinction law of \citet{2019Wang}. (bottom) IRTF/Spex spectrum of HOPS 154 (left) and Gemini/Flamingos 2 spectrum of GM Cha (right). In the case of HOPS 154 the upper panel shows the J, H and K spectrum of the source. In the bottom panel for both sources, only the K-band portion of the spectrum is shown, along with the location of typical emission/absorption features in YSOs.}
    \label{fig:hops154}
\end{figure*}

\begin{figure*}
	\resizebox{\columnwidth}{!}{\includegraphics[angle=0]{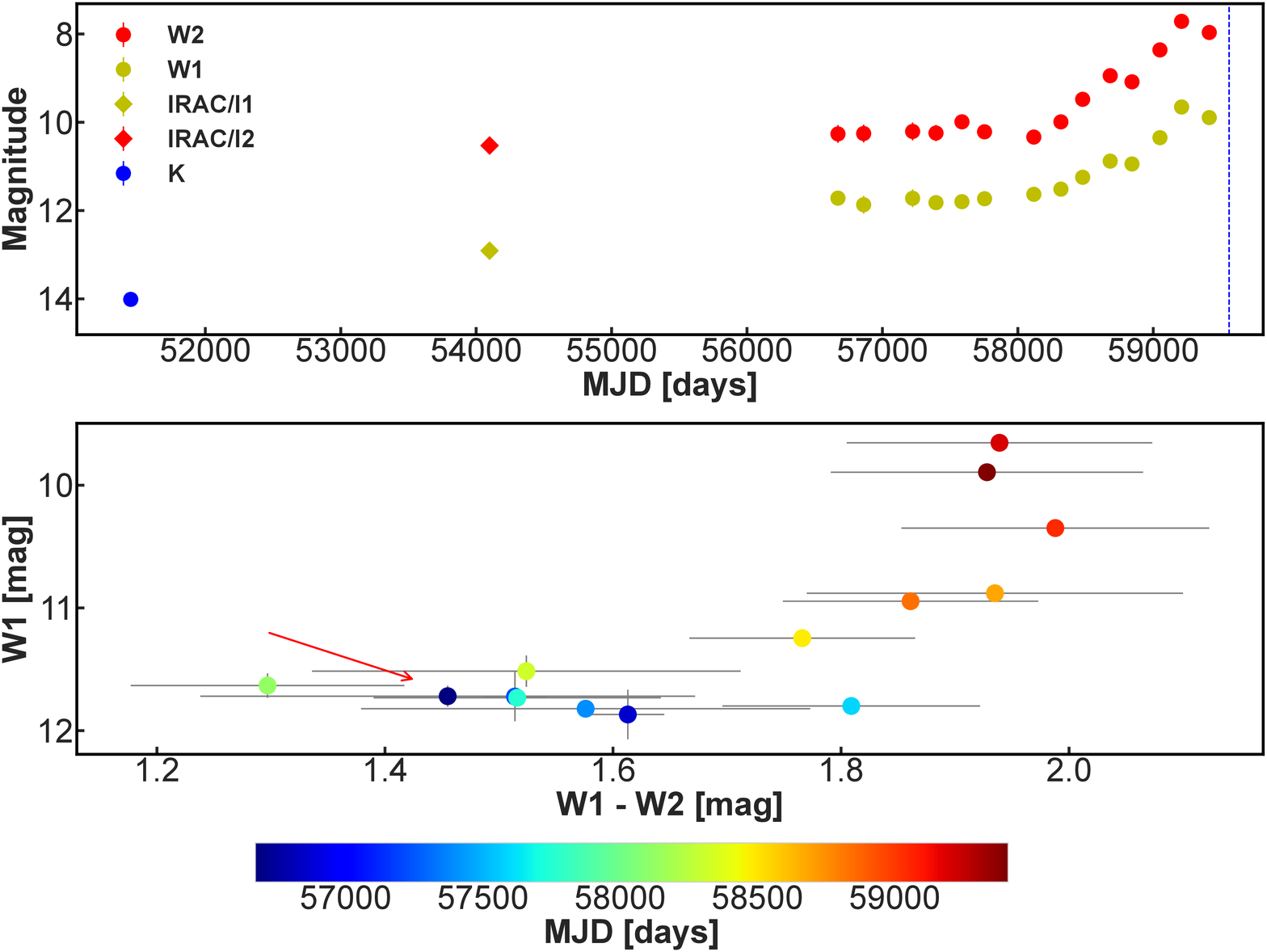}}
	\resizebox{\columnwidth}{!}{\includegraphics[angle=0]{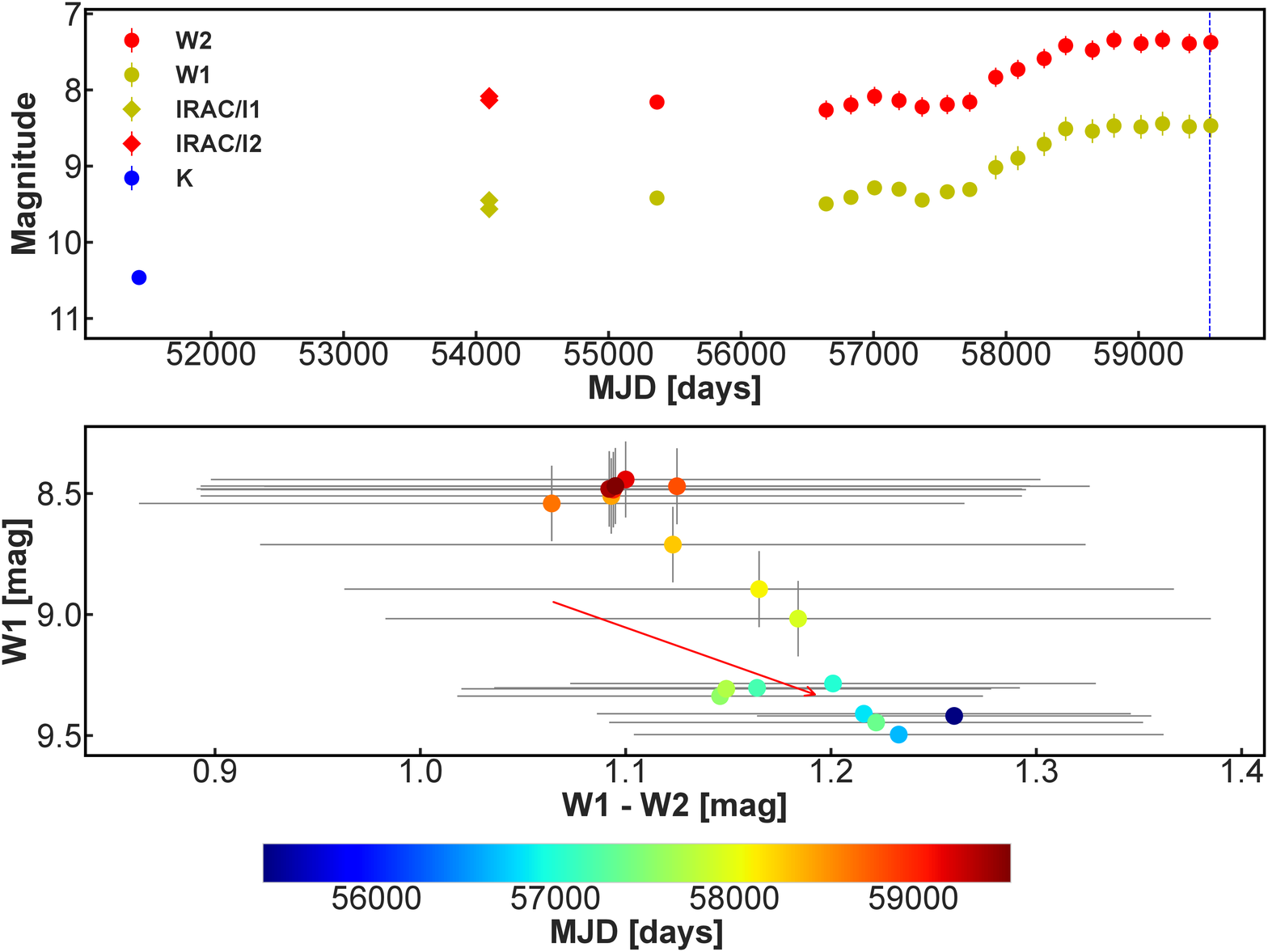}}\\
	\resizebox{\columnwidth}{!}{\includegraphics[angle=0]{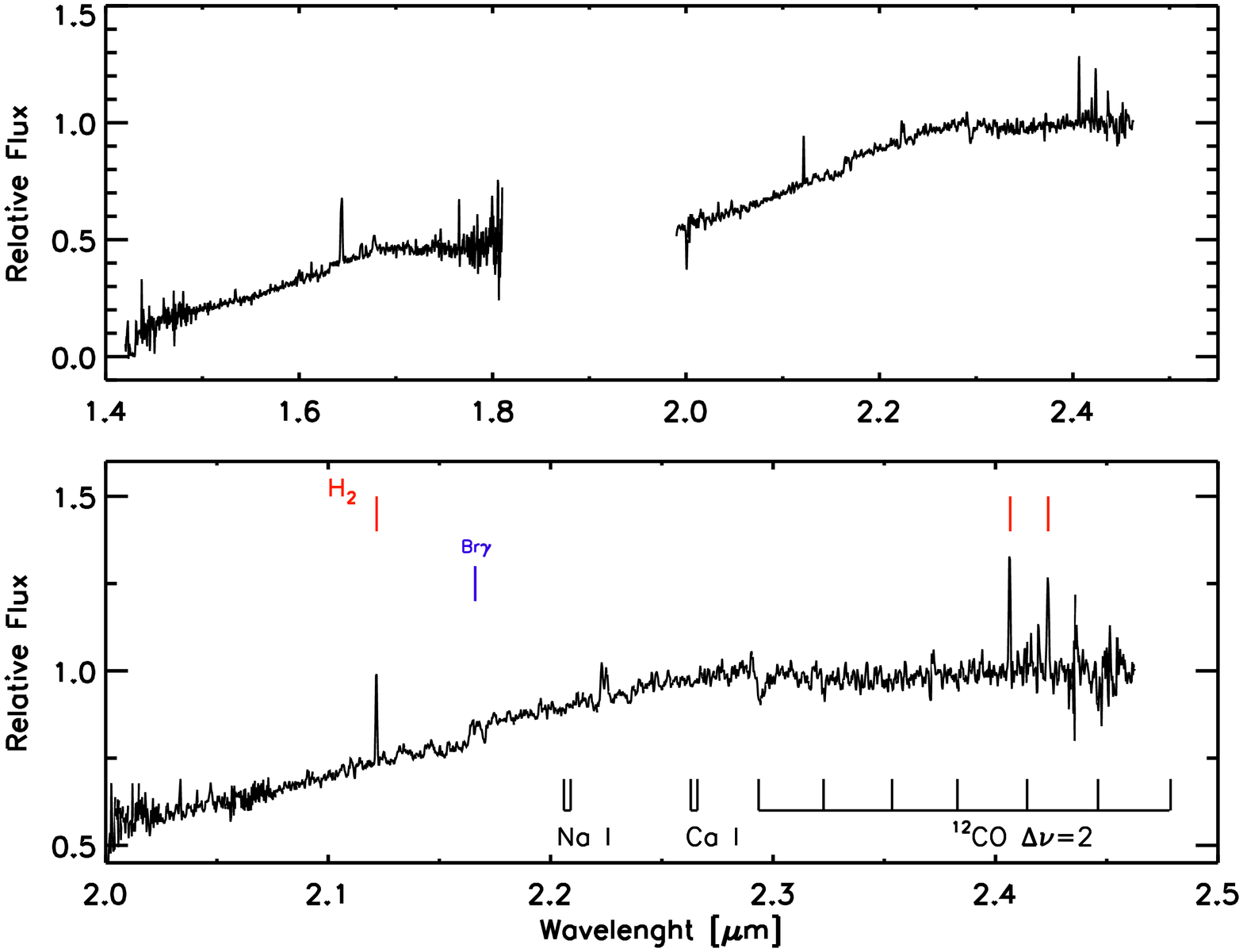}}
	\resizebox{\columnwidth}{!}{\includegraphics[angle=0]{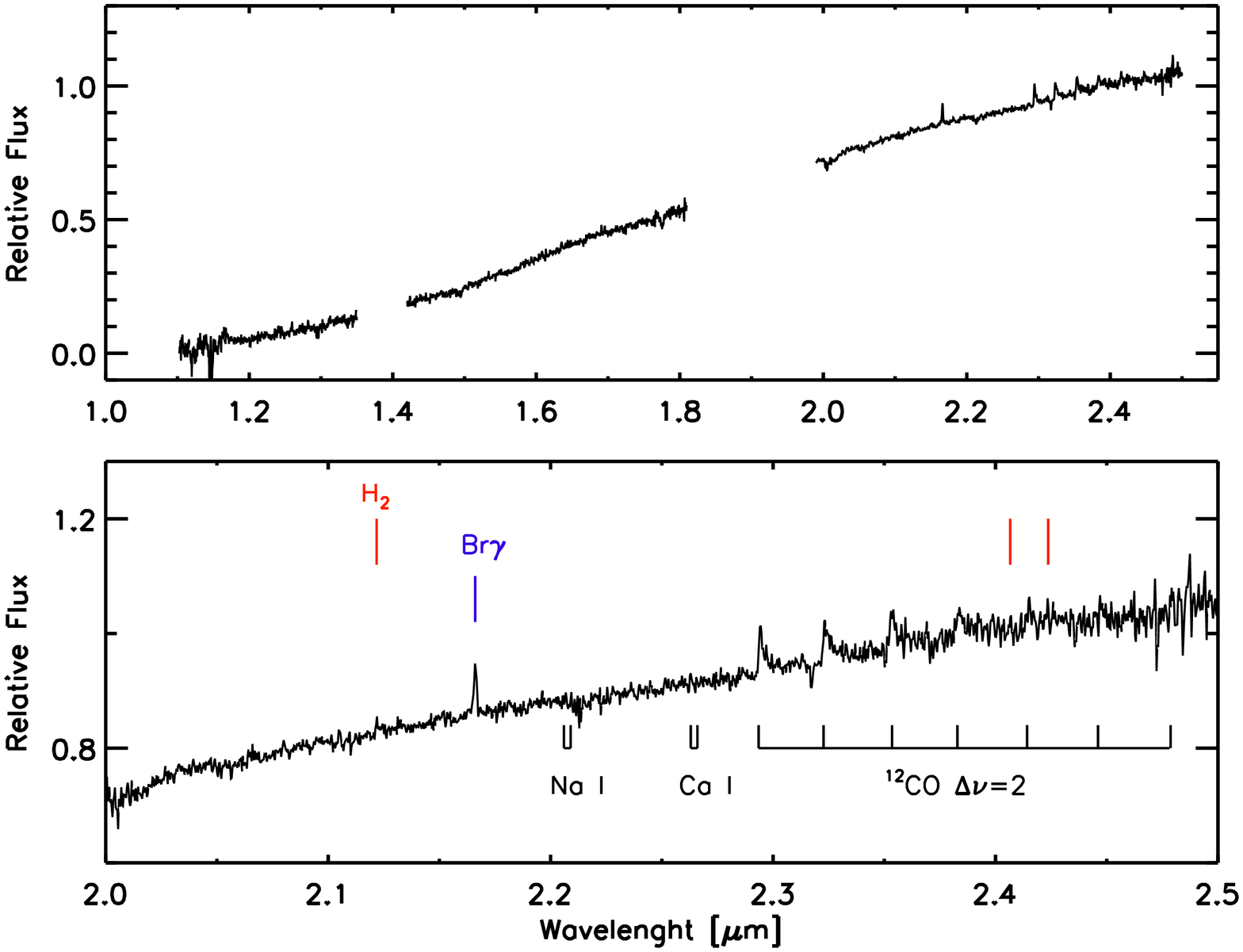}}
	 \caption{(top) K-band (blue), 3.4 $\mu$m (yellow) and 4.6 $\mu$m (red) light curve of 2MASS J21013280$+$6811204 (left) and 2MASS J21533472$+$4720439 (right). The date of spectroscopic observations is marked by a dashed blue line. (middle) $W1$ vs $W1-W2$ colour-magnitude diagram for 2MASS J21013280$+$6811204 (left) and 2MASS J21533472$+$4720439 (right). The plots only include data obtained by the {\it WISE} telescope. The red arrow marks the reddening line for A$_{V}=20$~mag, using the extinction law of \citet{2019Wang}. (bottom) Palomar/TripleSpec spectrum of 2MASS J21013280$+$6811204 (left) and IRTF/Spex spectrum of S2MASS J21533472$+$4720439 (right). The upper panel shows the J, H and K spectra of the sources. In the bottom panel, only the K-band portion of the spectrum is shown, along with the location of typical emission/absorption features in YSOs.}
    \label{fig:4990}
\end{figure*}

\begin{figure*}
	\resizebox{\columnwidth}{!}{\includegraphics[angle=0]{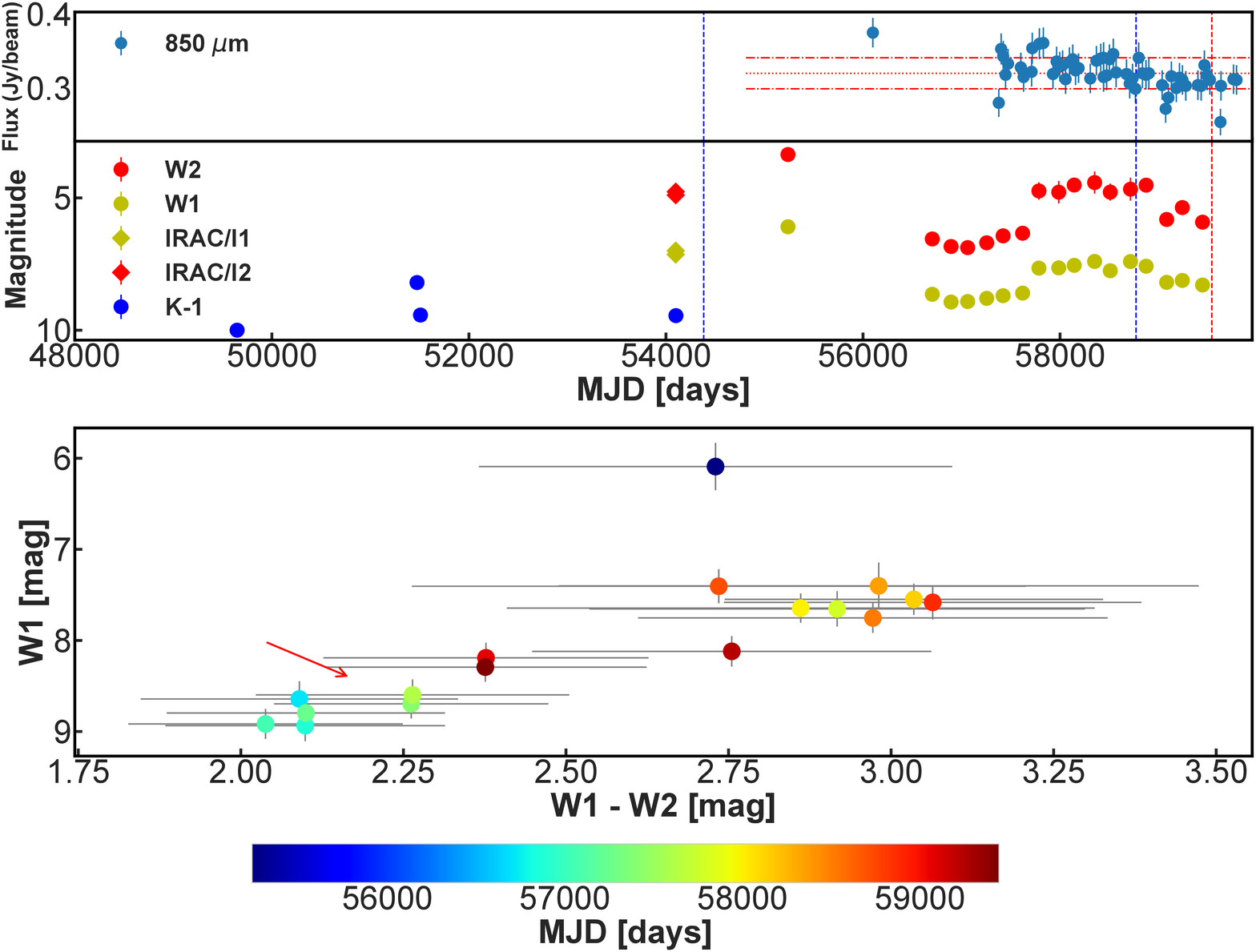}}
	\resizebox{\columnwidth}{!}{\includegraphics[angle=0]{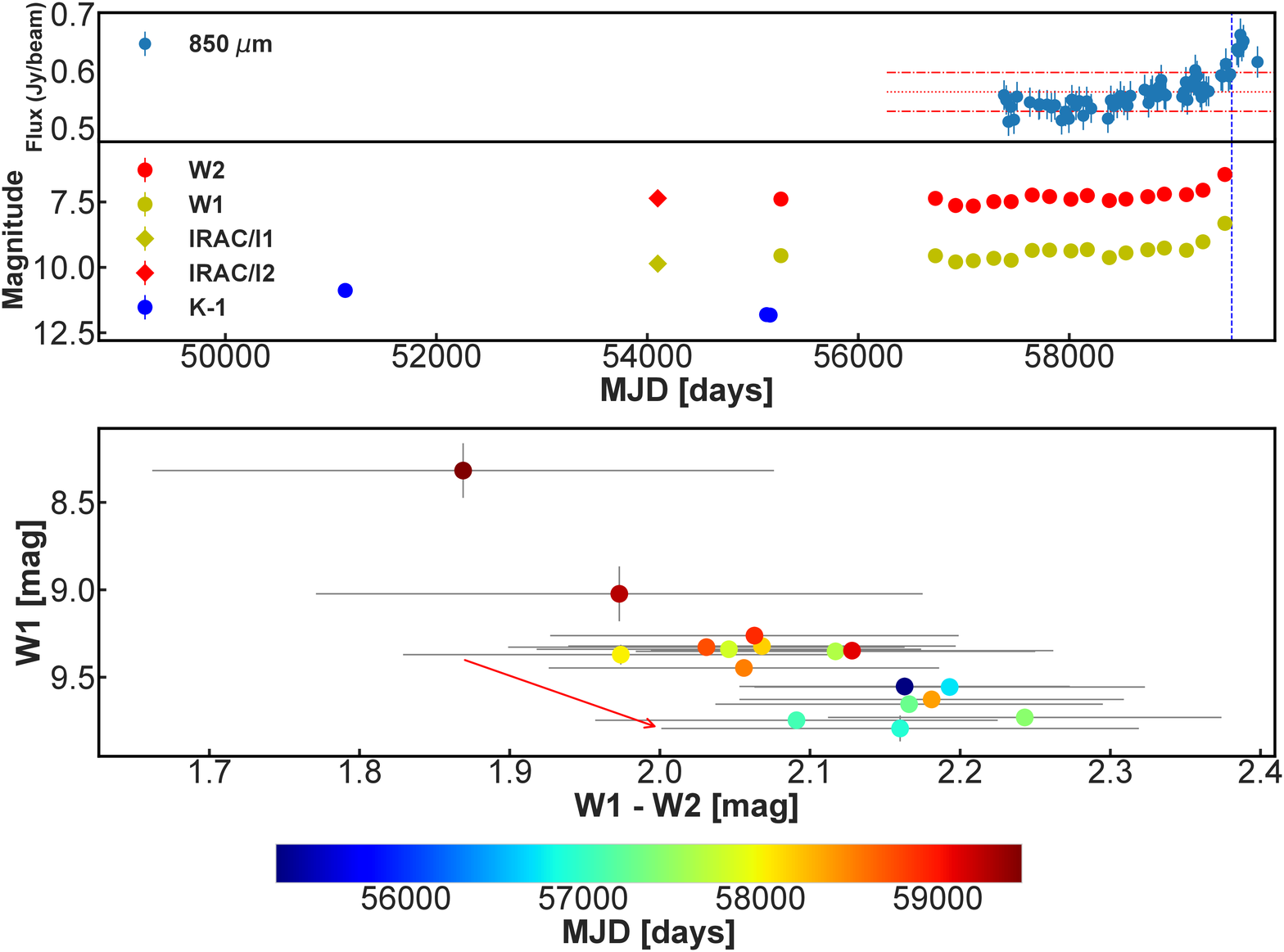}}\\
	\resizebox{\columnwidth}{!}{\includegraphics[angle=0]{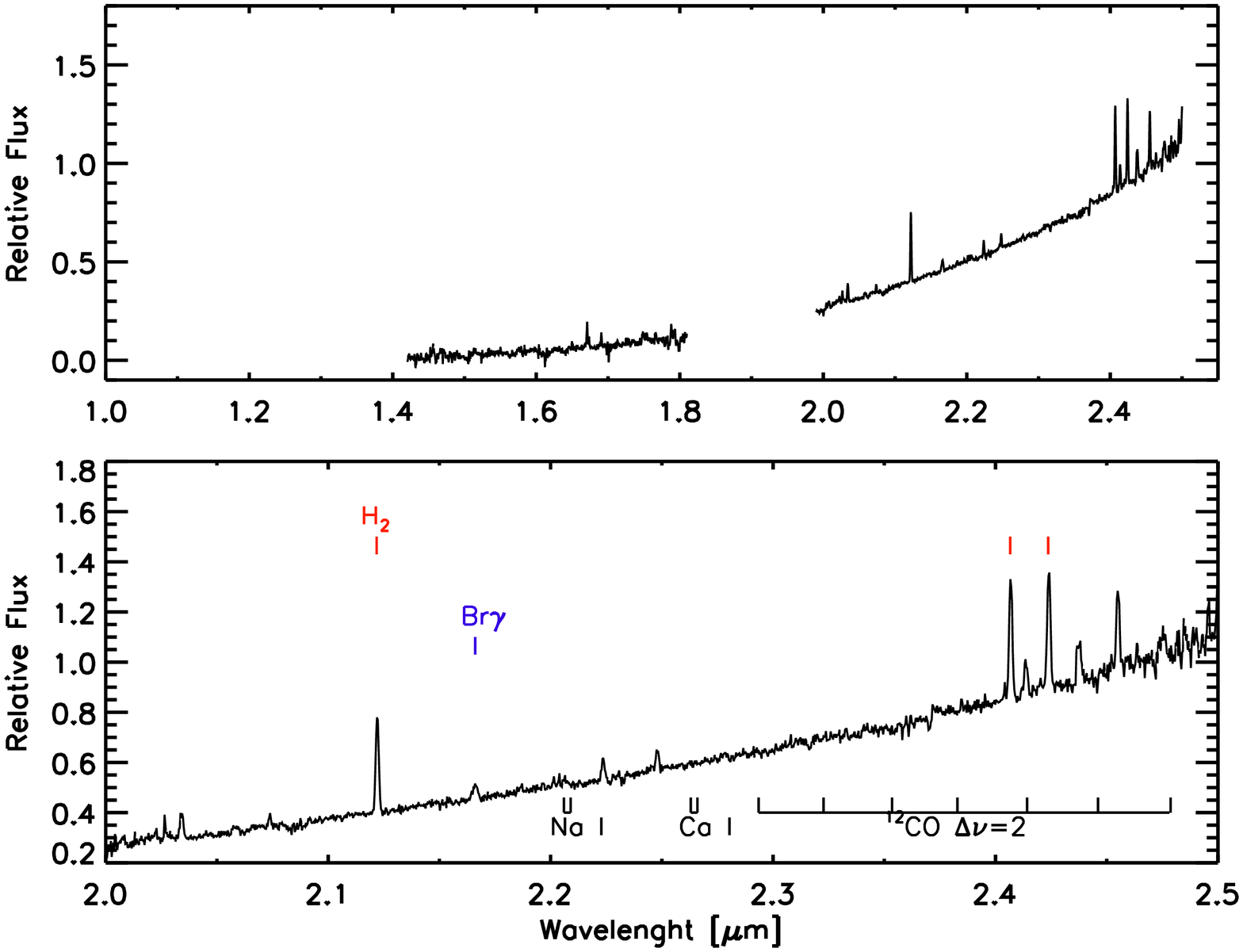}}
	\resizebox{\columnwidth}{!}{\includegraphics[angle=0]{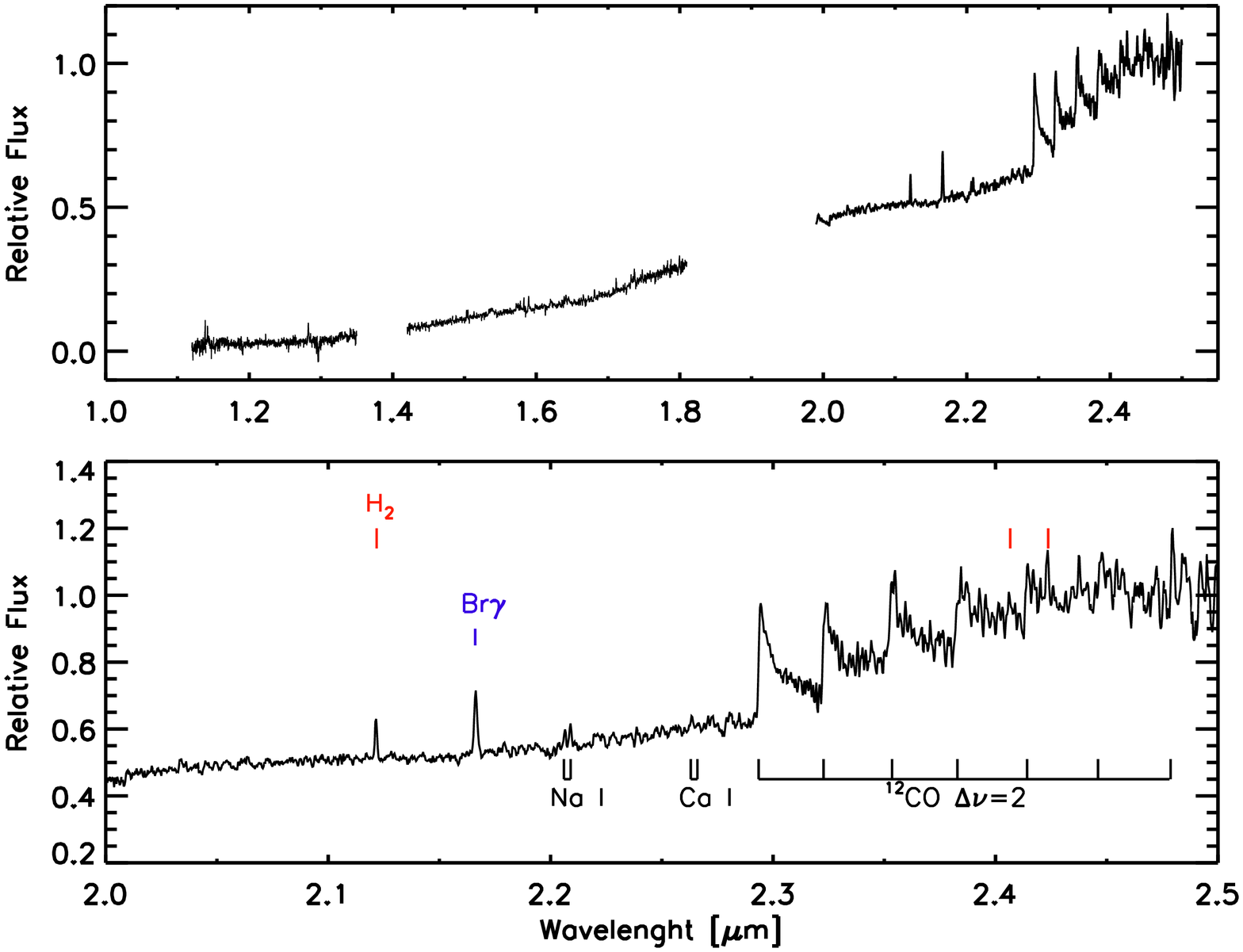}}
	 \caption{(top) K-band (blue), 3.4 $\mu$m (yellow), 4.6 $\mu$m (red) and 850 $\mu$m (light blue) light curves of [LAL96] 213 (left) and HOPS 315 (right). The date of spectroscopic observations is marked by a dashed blue line. The date of spectroscopic observations from this work is marked by a dashed red line. For YSO [LAL96] 213, the date of previous spectroscopic observations by \citet{2010Connelley} and \citet{2021Fiorellino} are marked by red dashed lines. (middle) $W1$ vs $W1-W2$ colour-magnitude diagram for [LAL96] 213 (left) and HOPS 315 (right). The plots only include data obtained by the {\it WISE} telescope. The red arrow marks the reddening line for A$_{V}=20$~mag, using the extinction law of \citet{2019Wang}. (bottom) IRTF/Spex spectrum of [LAL96] 213 (left) and HOPS 315 (right). The upper panel shows the J, H and K spectra of the sources. In the bottom panel, only the K-band portion of the spectrum is shown, along with the location of typical emission/absorption features in YSOs.}
    \label{fig:hops315}
\end{figure*}

\begin{figure}
	\resizebox{\columnwidth}{!}{\includegraphics[angle=0]{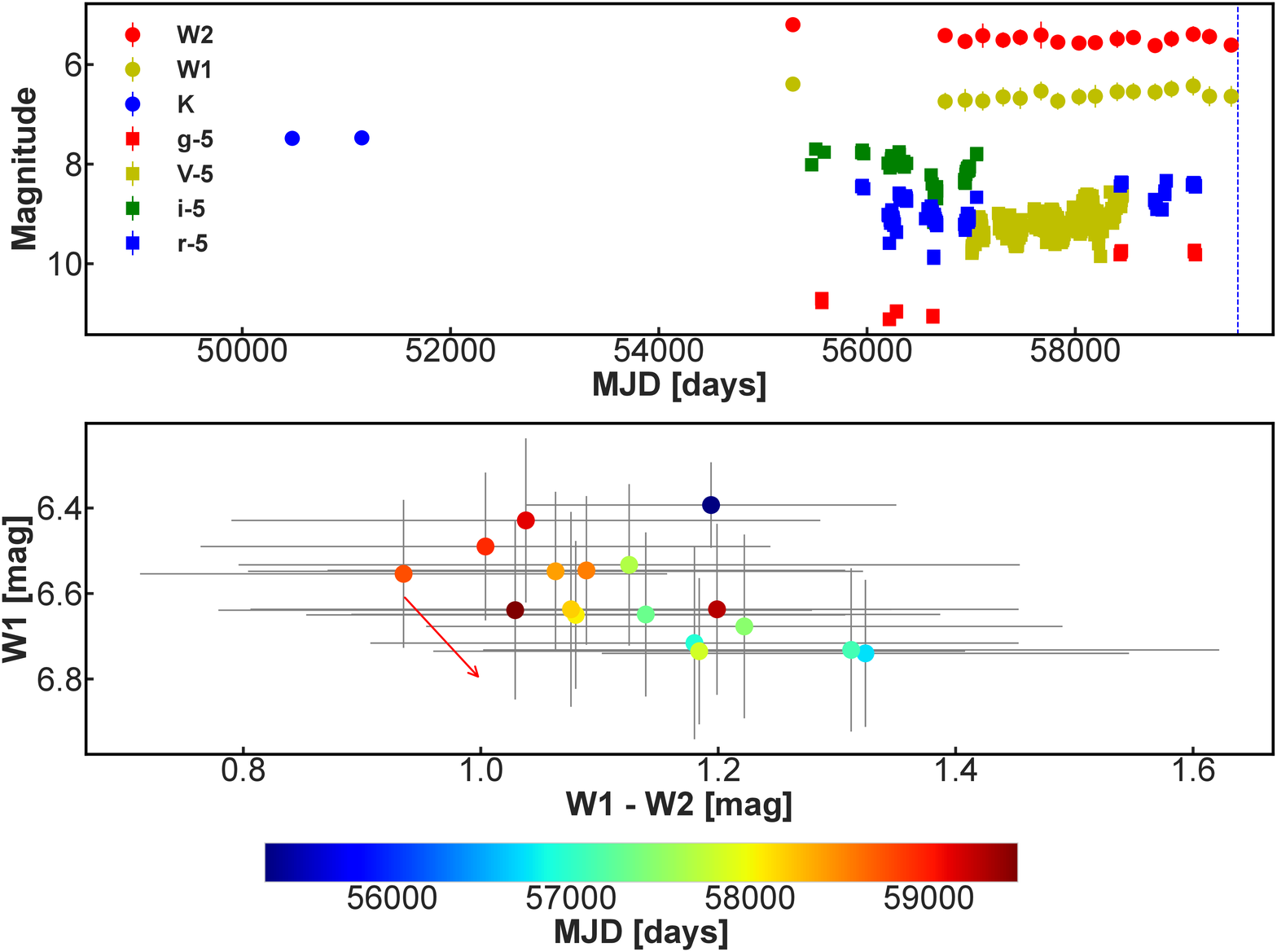}}\\
	\resizebox{\columnwidth}{!}{\includegraphics[angle=0]{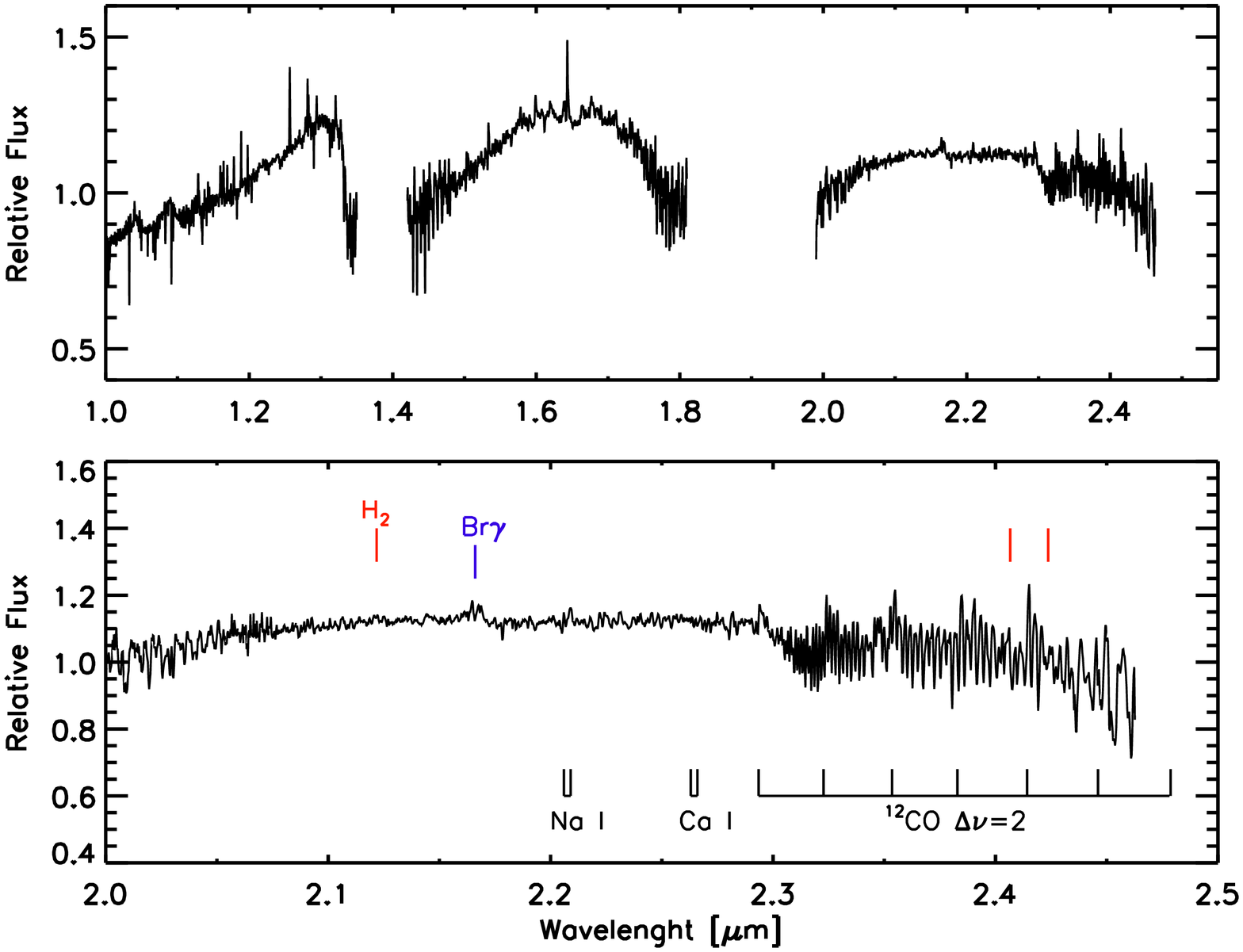}}
	 \caption{(top) sloan $g$ (red square), sloan $r$ (blue square), sloan $i$ (green square), V-band (yellow square), K-band (blue circle), 3.4 $\mu$m (yellow circle) and 4.6 $\mu$m (red circle ) light curve. An offset of 5 magnitudes has been applied to the optical data. The date of spectroscopic observations is marked by a dashed blue line. (middle) $W1$ vs $W1-W2$ colour-magnitude diagram of the same YSO. The plot only includes data obtained by the {\it WISE} telescope. The red arrow marks the reddening line for A$_{V}=5$~mag, using the extinction law of \citet{2019Wang} (bottom) Palomar/TripleSpec spectrum of V565 Mon. The upper panel shows the J, H and K spectrum of the source. In the bottom panel, only the K-band portion of the spectrum is shown, along with the location of typical emission/absorption features in YSOs.}
    \label{fig:V565}
\end{figure}


\bsp	
\label{lastpage}
\end{document}